\newcommand{\Diff}{\mbox{$ {\cal D}$}}

\documentstyle[aps,prb]{revtex}
\begin{document}
\newcounter{Table1}
\setcounter{Table1}{1}

\twocolumn[\hsize\textwidth\columnwidth\hsize\csname@twocolumnfalse%
\endcsname

\title{
Mesoscopic and microscopic dipole clusters: Structure and phase
transitions
}

\author{A.I. Belousov and Yu.E. Lozovik\cite{E}}

\address{
Laboratory of Nanophysics,
Institute of Spectroscopy, Russian  Academy  of  Sciences,\\
Troitsk, 142092, Moscow region, Russia}

\maketitle

\begin{abstract}
Two dimensional (2D) classical system of dipole particles
confined by a quadratic potential is studied.
This system can be used as a model for rare electrons in
semiconductor structures near a metal electrode, indirect
excitons in coupled quantum dots etc.
For clusters of $N \le 80$ particles ground state configurations
and appropriate eigenfrequencies and eigenvectors for the normal
modes are found.
Monte Carlo and molecular dynamic methods are used
to study in detail the order - disorder transition (the "melting"
of clusters).
In mesoscopic clusters ($N < 37$) there is a hierarchy of
transitions:
at lower temperatures an intershell orientational disordering
of pairs of shells takes place;
at higher temperatures the intershell diffusion sets in and
the shell structure disappears.
In "macroscopic" clusters ($N > 37$) an orientational
"melting" of only the outer shell is possible.
The most stable clusters (having both maximal lowest nonzero
eigenfrequencies
and maximal temperatures of total melting) are that of completed
crystal shells which are concentric groups of nodes of
2D hexagonal lattice with a number of nodes placed in the center
of them.
The picture of disorderung in clusters
is compared with that in infinite 2D dipole system.
The study of the radial diffusion constant,
the structure factor, the local minima distribution and other
quantities shows that the melting temperature is a nonmonotonic
function of the number of particles in the system.
The dynamical equilibrium between "solidlike" and
"orientationally disordered" forms of clusters is considered.


\end{abstract}

]

\section{Introduction}
\label{introduction}
Properties of a set of physical systems can, under certain
conditions, be studied with the help of the model system of
interacting dipoles.
Particularly, it is a dipole - dipole interaction that is of
main importance when small dielectric particles on a surface
of electrolyte \cite{Pilranski}
or monolayers of adsorbed atoms \cite{Bolsov} are considered.
Another class of the systems in which the most
part of interesting properties is due to a dipole character of
interaction is the class of electron systems
in semiconductors of small size located
in the vicinity of a metal electrode electrostatic
image forces can play an essential role. This role can become
crutial, for example,
for semiconductors with a small width of the forbidden band,
for which image forces can induce a semiconductor - metal
transition. \cite{Agran}
Indirect excitons with spatially separated electrones and
holes in coupled quantum wells also interact by a dipole law,
leading to the crystallization at intermediate
concentrations. \cite{LozExciton,McDonald}

Properties of {\it infinite} two dimensional
dipole systems have been intensively studied
both theoretically \cite{2DPhonon} and with the help of
computer simulations \cite{2DDip}
(see also Ref. [\onlinecite{2DDipE}]).
But the study of mesoscopic systems of interacting
dipoles (dipole clusters) is also of a particular interest.
Recent advances in the microlithography and one - electronics
revived markedly an interest to systems of small number of
particles, interesting with the strong structural sensitivity
to the number of particles and unusual rearrangements with
increasing temperature.
Particularly in recent experiments of Ashoori
et al. \cite{West}
the charging of a quantum dot was studied by the single
electron capacitance spectroscopy method.
One of the most intriguing results of these experiments was
the observation of an unusual dependence of the chemical potential
on the particle number $N$, when the total energy of the dot as a
function of $N$ had a quasi - periodic component of a universal
shape (i.e. the electron addition are group into bunches).
This dependence was shown to be caused by geometric effects
associated with a packing of the triangular lattice into the
circular confinement. \cite{Koulakov}
It is obvious that if the dot of a large enough size is
situated near a single metal gate the electrostatic image forces
will play an important role and
the long-range electron - electron Coulomb interaction potential
will transform to
the short-ranged dipole - dipole one between "composite" particles
"electron $+$ image charge".
A plane microtrap of cooled atoms with
(normal to the plane) dipole momenta induced by
static electric field or by an
electromagnetic standing wave could be another physical example
of a dipole cluster.
The value of the dipole momenta of trapped atoms
can be controlled by varying the magnitude of the field or
by tuning the field frequency.
Yet another interesting example of a dipole cluster
is a system of indirect excitons in vertically coupled semiconductor
quantum dots (see e.g. Refs. [\onlinecite{Butov,Dots}]).
Under certain conditions
the system of indirect excitons can be considered as a
classical dipole cluster in the parabolic confinement (see below).
Such double dots systems are within the scope of
current fabrication technology.

The consideration of different properties of
dipole, Coulomb and logarithmic clusters
provides also a possibility to study the role of the interaction
potential in the shell structure of ground state configurations
(at $T = 0$) and in the hierarchy of melting temperatures.
One of the most interesting tasks is the
search for "magic", most stable clusters.
By now the structure of magic number clusters has been most
extensively studied by the example of Lennard - Jones ($LJ$)
systems that correspond, e.g., to noble gases atomic clusters.
\cite{Berry}$^{-}$\cite{Berry_1}
Of some interest is the question: does the set of distinctive
properties of "magic" clusters correlate with the particular
form of an interaction potential?
It is important to realize the manner in which changing the
type of interaction potential from two - parameter in $LJ$ system
to repelling one - parameter in a dipole system may
reflect on the properties of appropriate "magic" clusters,
on the character of disordering effects and on the possibility
of coexistence phenomena.

In this paper we study in detail the ground state structure
of 2D dipole clusters consisting of a finite number of particles
and the picture of their melting with increasing temperature.
It turnes out that on the basis of the character of disordering
("melting") phenomena dipole clusters can be divided
into "mesoscopic" and "macroscopic" groups.
In "mesoscopic" clusters of small number of particles ($N < 37$)
there are {\it two} stages of disordering:
{\it orientational intershell disordering} of different pairs of
shells and, at greater temperatures,
{\it radial} disordering when particles begin to
interchange between shells.
The spectral analysis of ground state configurations of small
clusters shows that the value of the minimal frequency
of normal vibrations correlates with
characteristic temperatures of orientational disordering.
Increasing of the temperature of large particle number
clusters ($N > 37$) leads to radial disordering and to
disruption of the shell
structure at a temperature which is a function of the distance
from the center of the system.
An analysis of a behavior of different thermodynamic
properties, when the nonuniformity of the density
is taken into account, enables us to make a conclusion about the
strong dependence of the total melting temperature on a cluster
position in a periodic Mendeleev - type table (i.e. on the
distribution of particles over crystal shells of different
symmetry).
As an intermediate
between mesoscopic and macroscopic systems
the cluster of $37$ particles is considered.
The most interesting feature
of this cluster is the presence of the temperature region of
coexistence of "solidlike" and orientationally disordered forms.
This behavior of the system with increasing temperature is
much alike the picture of structural transitions in magic
number atomic clusters.

The paper is organized as follows.
In Sec.~\ref{methods} we describe the model
and briefly outline the methods that where used to find
global minima configurations and to calculate different
thermodynamic properties.
Sec.~\ref{ground_config} is devoted to the description and
discussion of ground state configurations.
We consider the "faceting" of clusters,
the eigenvectors and eigenfrequencies for the normal modes.
In Sec.~\ref{transitions} we present the results from Monte Carlo
(MC) and molecular dynamic (MD) simulations at different
temperatures and system sizes. We consider two cases: "mesoscopic"
($N < 37$, see Sec.~\ref{small}) and "macroscopic"
($N > 37$, see Sec.~\ref{large}) clusters.
Separately, as an intermediate case, in Sec.~\ref{D_37}
we study the cluster of $N=37$ particles.
Our conclusions are presented in Sec.~\ref{conclusions}.

\section{The model. Numerical simulation}
\label{methods}

The exposition below concerns a cluster of indirect
excitons in double semiconductor dots,
but the analysis can be obviously extended to any other
above - mentioned dipole system.
Let us consider two vertically coupled 2D semiconductor dots of a
characteristic linear size $L$ that are spaced at $h$ apart.
Under laser pulse on the forbidden band frequency,
a formation of indirect excitons that are bounded states of
electrons and holes belonging to different dots is possible
(see. works [\onlinecite{Butov,Dots}] and references therein).
In this work we are interested in the case that the
exciton radius $a(h)$
is much smaller of the mean distance $L / \sqrt{N}$ between them;
$a(h) \sim a_0^{*}$ at the interdot separation $h < a_0^*$
and $a(h) \sim {a_0^*}^{1/4} h^{3/4}$ at $h \gg a_0^*$ where
$a_0^* = \hbar^2 \kappa / (4 m^* e^2)$ is  effective Bohr
radius of 2D exciton in a media with dielectric constant
$\kappa$, $N$ stands for the (mean) number of excitons in
the system.
In addition we suppose that quantum effects in the cluster
can be neglected in view of the condition
$k_b T \gg \hbar^2 N / (m^* L^2)$.

Provided that conditions discussed above are fulfilled, excitons in
quantum dots can be considered as 2D classical system of particles
with unidirected dipole momenta $d \approx 2e h / \kappa$.
A confinement potential can be chosen in a parabolic form with
strength $\alpha$.
The Hamiltonian for such a system have the form
$$
H = \sum\limits_{i<j} \frac{d^2}{|{\bf r}_{i} - {\bf r}_{j}|^3} +
\alpha \sum_{i=1}^N |{\bf r}_{i}|^2
$$
The Hamiltonian can be written in a dimensionless form if we
express the coordinates and energy in the following units:
$r_0 = d^{2/5} / \alpha^{1/5}$, $E_0 = \alpha r_0^2$.
In such units the Hamiltonian becomes
\begin{eqnarray}
H = \sum\limits_{i<j} \frac{1}{|{\bf r}_{i} - {\bf r}_{j}|^3} +
\sum_{i=1}^N |{\bf r}_{i}|^2
\label{E}
\end{eqnarray}
From here on all the results will be given in the units
introduced above.
When considering thermodynamic properties of
the system we will use the dimensionless temperature
$T = k_b T / \alpha r_0^2$.

To find out ground state configurations of (\ref{E}) we applied
the following methods:

1) {\it Classical simulated annealing} method (CSA).  \cite{CSA}

2) {\it Combined Monte Carlo} + {\it gradient search"} method.

All of the results presented below
(see Table~\Roman{Table1}) were
independently obtained with the help of these algorithms in order to
improve the reliability of results. Of course, none of existing
algorithms of searching for the global minimum of multidimensional
function can
guarantee that a configuration obtained is the global minimum one.
In order to overcome this difficulty we considered as many as 200
random initial configurations. This approach made it possible to
study local minima as well as appropriate regions of catchment
("relative weights" of local minima).

In view of intrinsic statistical nature of the {\it CSA} method,
the problem of localization of the system in local minima is solved
in this approach much
more easy than in different gradient methods.
When using CSA,
the system is modelled at some artificially introduced
temperature $T(t)$ which is gradually decreased with the time
$t$ of experiment.
This make it possible to simulate the thermal noise and
delocalize a system from metastable states.
By starting from sufficiently high temperature of $T(t=0) = 10$,
at each $t$ we performed $\sim$200 MC steps per particle,
whereupon the temperature was decreased as $T(t+1) = 0.98 T(t)$.
The trial move $x \to x + \delta x$
at each MC step was generated in accordance with Gaussian
probability distribution
$\Pi(\delta x) \sim \exp{(-\delta x^2 / T(t))}$.
This move was accepted with the probability
$p = \min\{ 1, \exp{(-\triangle E / T(t))} \}$.
To gain in the accuracy of results several hundreds of
conjugate gradient
steps were performed at the last stage of the algorithm.

It turned out that CSA is not as fast the algorithm as the
{\it combined MC + gradient search} method.
A search for minima within the last approach consists of
repeated (up to $10^4$ combined moves) applying of the
following methods:

1) Classical random search \cite{Mandelstam} when the trial
move $x \to x + \delta x$ is accepted if it reduces the energy.
The maximal size of the trial move $\max \{ |\delta x| \}$
was chosen automatically to assure the acceptance probability
of a new configuration of $0.1$.

2) Gradient search method
$ x \to x - \gamma \nabla H, \; \gamma \approx 0.01$.

3) Ravine method.\cite{Kalitkin}
In this method the direction $ l =  x_m - x_1$
of the most probable displacement of the system that moves
in a long narrow ravine is predicted on the basis of a series
$\{ x_1, x_2,... x_m \}$
of previously performed gradient search steps.
A number of trial moves in this direction
$ x \to 0.5( x_1 + x_2) + l \delta$
completes the method.
The maximal size of these moves $\max \{ |\delta| \}$
was adjusted in
such a way as to achieve the acceptance probability of $\sim 0.1$.
We found that the throughoutput of the combined method was maximal
when moves of above - mentioned types where performed in the ratio
of $N_1 : N_2 : N_3 = 100 : 10 : 2$.

When studying the thermodynamic properties of the system, we used
the Metropolis algorithm. \cite{Metrop}
For the system of small number of particles ($N < 40$) and at
sufficiently low temperatures ($T < 0.02$), when shells are well
defined, we found it very efficient to perform collective MC moves
of different shells. With the multigrid approach \cite{MultiGrid}
for shell $s$ of $N_s$ particles one can apply the following types
of collective moves with wave vector $k_s$:

1) angular perturbations such that angles $\varphi_{i_s}$ of
particles belonging to shell $s$ vary as
\begin{eqnarray}
\varphi_{i_s} \to \varphi_{i_s} +
\xi \delta_{\varphi}(k;s) \cos{(2\pi k i_s / N_s)}
\label{d_phi}
\end{eqnarray}
with $\xi$ being the random variable uniform in $[-1,1)$.

2) radial perturbations of shell $s$:
\begin{eqnarray}
r_{i_s} \to r_{i_s} +
\xi \delta_r(k;s) \cos{(2\pi k i_s / N_s)}
\label{d_r}
\end{eqnarray}

It is obvious that the case $k_s = 0$ corresponds to the
symmetrical "breathing" of shell $s$ when the radial perturbation
2) is concidered
and to the rotation of the shell as a whole when one performs
the global move of the first type. We found that such rotation
of different shells
was very important when studying the phenomena of
intershell orientational disordering which take place in
mesoscopic systems at low temperatures
(the temperatures of orientational "melting" can be many orders
less than temperatures of full disordering and shells destruction).
The use of the collective moves
described above enabled us to increase an efficiency of
calculations, that is inversely proportional to the time
needed to obtain a thermodynamic average with a given accuracy,
about a five times.

Parameters $\delta_{\varphi}$ and $\delta_{r}$
(in Equations (\ref{d_phi}),(\ref{d_r})),
as well as the maximum size of the Metropolis trial move, were
adjusted in such a way as to hold the acceptance probability
$p_a \approx 0.4$. To do this
we averaged the acceptance probability $\left< p \right>$
of trial moves of
different types
($\left< p_r \right>,\; \left< p_{\varphi} \right>$)
in $\sim 50$ MC steps per particle
followed by the scaling
$\delta_{\varphi}(k;s) \to \delta_{\varphi}(k;s) (1-p_a) /
\left< p_{\varphi} \right>$ (and analogous for $\delta_{r}$).

To study dynamical characteristics of the system we applied the
molecular dynamic  simulation (both isokinetical and microcanonical
methods where used).
The main portion of the results presented
below were obtained with the help of the fourth - order
Runge - Kutta scheme.
Equations of motion were integrated in $\sim 10^4$ MD
steps of size $\tau \le 0.01 \tau_0$, where
$\tau_0 = \sqrt{\alpha / m^*}$ determines the time scale
in the system.

\section{Ground State Configurations}
\label{ground_config}

Previous studies of Coulomb \cite{Mandelstam,Bedanov}
and logarithmic \cite{Calinon}$^{,}$\cite{Log,Rakoch_log}
clusters have shown that it is suitable to classify these
finite systems in accordance with their shell structure.
An analysis of
shell structures for different number of particles $N$
enables one to consider the system as belonging to some
period of a Mendeleev - type table.
This table can be viewed as a classical equivalent
to the well - known Periodic Table of elements.
By studying regularities in this table the
Coulomb clusters with a constant density of particles in
each shell have been explored.
This systems have been called "magic" Coulomb clusters.
As examples of such systems clusters of $18$
$(1,6,11)$, $30$ $(5,10,15)$ and of $57$ $(1,6,11,17,22)$
particles have been proposed.
The existence of magic clusters particularly
manifests itself as cusps at a plot of excess energy
$\epsilon(N) = E(N)/N$ as a function of the number of particles.
It have also been shown that these systems are more stable
against intershell rotation.

Interaction between particles in a dipole cluster is
more short - ranged
with respect to that in Coulomb or logarithmic systems.
Therefore it is possible to trace the role of the range of
interparticle potential in the structure of 2D clusters,
in the character of the formation of the Periodic Table
and the type of shells (see also Ref. [\onlinecite{Koulakov}]).
A difference between structures of dipole (D) and Coulomb clusters
can be seen even at $N=10$ (the system $D_{10}$ has the structure
$D_{10}(3,7)$ as distinct from $(2,8)$ for Coulomb
system. \cite{Rakoch_dip})
The number of differences grows rapidly with increasing the number
of particles. To explore tendencies in the
process of cluster shells developing,
let us consider the shell configurations of 2D dipole clusters
that are presented in Table~\Roman{Table1}.
The table shows that
the basis for most configurations is provided by different parts
of 2D hexagonal lattice.
It should be noted that this observation
is not specific to clusters with short - ranged interaction only.
As was argued by Koulakov and Shklovskii [\onlinecite{Koulakov}],
only a narrow ring adjacent to the perimeter of sufficiently
large Coulomb clusters is concentric to it, the rest
of the cluster is filled with an almost perfect crystal.

When describing and analyzing the properties of such configurations
we found it suitable to introduce into consideration
the "crystal shells" $Cr_c$ that are concentric groups of nodes of
ideal 2D crystal with $c$ nodes placed in the center of these groups
(see Fig.~1).
Obviously, in view of the axial symmetry of the confinement potential,
we can concentrate on a finite number of the most symmetrical crystal
shells.

By the number of particles in the center of the system,
the crystal shells can be divided into the following groups:
$Cr_{1}$, $Cr_{2}$, $Cr_{3}$, $Cr_{4}$.
Fig.~1 explains our definition. The number of particles $N_s$
that belong to crystal shell $s$ (crystal row)
of type $Cr_c$ is $N_s = c + 6(s-1)$.
With the help of the crystal shell concept
an analysis of the results presented in Table~\Roman{Table1} shows
that a formation of a "Periodic Table of elements" is done in
accordance with the following {\it tendencies}:
\begin{enumerate}
\item The maximum number of {\it crystal shells} is filled.
\item The number of particles on the last two shells
tends to be equal.
\end{enumerate}
To illustrate these tendencies we underline
the shells which are filled up (see Fig.~2 and Table~\Roman{Table1}
where types of basic crystal shells are also shown).

An addition of particles to the cluster leads to the completing of
crystal shells of some type, followed by the structure rearrangement
(see Fig.~2(c)-2(d)) after which a crystal shells of different type
begins to fill.
From Table~\Roman{Table1} one can see that in some cases it is more
advantageous
to departure from the order $Cr_1 \to Cr_2 \to Cr_3 \to Cr_4$
in which different types of crystal shells appear.
Examples of such deviations (see $D_{56}-D_{61}$ and $D_{71}-D_{73}$)
show that it can be profitable to reduce the number of particles
on first shells (in the center of the system) and equalize them
on the last ones. This observation underlines the importance
of the requirement that the last two shells of the cluster have equal
number of particles.

It is to be noted that for some systems the choice of the basis
crystal group is ambiguous. The most spectacular examples of this
ambiguity are clusters that we assign to $Cr_2$ crystal group
(such clusters are $D_{38},D_{59},D_{61},D_{62}$, see
Table~\Roman{Table1} and
Fig.~3). Of course, one can consider these systems as having
a number of partially completed crystal shells of $Cr_4$
type (the  "body" of a cluster) while the other particles
locate on the surface of this "body". This approach is illustrated
in Fig.~3 where two possible types of breakup
of cluster $D_{62}$ into shells is shown.
In choosing between these possibilities we have being guided
by the requirement the outer shell of the cluster to be
well - defined and {\it closed}.
For the system with a more short - ranged interaction potential
(e.g. with the exponential or screened Coulomb interaction
potential) it can be more natural to consider the evolution of the
cluster with $N$ as consisting of switching between {\it three}
types of the most symmetrical basic crystal shells,
namely $Cr_1, Cr_3, Cr_4$. \cite{Koulakov}
Seemingly, the question of what approach is more adequate can be
answered on the data of the thermodynamic and dynamic analysis.

Results presented in Table~\Roman{Table1} and above
consideration enable us to suppose
that for 2D dipole clusters a "period" of the Periodic Table
consists of clusters with partially completed crystal shells
of some symmetry (of some basic group).
A role of "magic" clusters will be played
by that with the {\it maximal number of completely filled
crystal shells}
and, in the case of clusters of large number of particles
($N > 40$),
with {\it equal number of particles on the two last shells}.
An analysis of Table~\Roman{Table1} suggests that dipole
clusters $D_{12},D_{14},D_{19},D_{36},D_{40},D_{51},D_{54},D_{55},
D_{62},D_{65}...$ can be considered as "magic" ones.
Plots of the first and the second "derivatives" of specific energy
$\epsilon(N)$ with respect to number of particles $N$ (see Fig.~4)
can serve as illustrations of this assumption.
The particle numbers at which cusps take place
correlate with the "magic" numbers.

Somewhat ambiguous is the arrangements of particles into shells in
the cluster $D_{37}$ (see Fig.~5).
We can see that one of the particles
is between the second and the third shell to form an
interstitial (it is analogous to the Frenkel defect in crystals).
Such classification of this particle is based on the Voronoi
analysis that shows that the central particle has six neighbors.
The corresponding configuration can be denoted by
$D_{37}(\underline{1,6},\tilde{1},13,16)$ where the tilde stands
for the interstitial.
There exist also other opportunities to describe the structure of
this cluster, for example as
$D_{37}(\underline{1,6,12},\tilde{2},16)$.
Our choice (see also Table~\Roman{Table1}
and Fig.~5(a)) is also proved itself
in the results of the analysis of the lowest local minima
configuration that
has well - defined shell structure: $D_{37}^{(1)}(1,7,13,16)$
This configuration is shown in Fig.~5(á).
It should be noted that the cluster involved
has a rather complex picture of disordering phenomena with
increasing temperature.
In the study of its thermodynamic properties
presented below we show that the system $D_{37}$ has properties
that are characteristic of both small ($N < 37$) and large
($N > 37$) clusters.

Another region of distinctions from the above tendencies in
particle distribution throughout shells
is a group of clusters $D_{70} - D_{75}$
that have a pentagonal shape.
It is possible that the size $\triangle N$ of the region of
non - hexagonal arrangement of $N$ - particle cluster facets
will be diminished with reducing the range of an interaction
energy.

To obtain further information about the character of shells
distribution in clusters of different number of particles we
applied the spectral analysis
of the ground state configurations (the spectral analysis of
Coulomb clusters was performed in the work [\onlinecite{Bedanov}]).
The main attention was given to the minimal nonzero
eigenfrequencies $\omega_{min}$ for the normal modes.
Fig.~6 shows $\omega_{min}$
as a function of the number of particles in the cluster.
It is worthwhile to note that magic number clusters have
maximal values of $\omega_{min}$. The correlation between
the value of $\omega_{min}$ and the extent to which crystal
shells are completed is also seen.

One of the most interesting features of clusters is the strong
dependence of their thermodynamic and dynamic properties on the
number of particles $N$. An investigation of this dependence
(see below) shows that dipole clusters can be divided into two
groups:
{\bf 1)} "mesoscopic" ("small") clusters, $N < 37$;
{\bf 2)} "macroscopic" ("large") clusters, $N > 37$.
The intershell orientational disordering phenomenon is distinctive
of mesoscopic clusters and does not take place in dipole clusters
of a large number of particles.
But differences between large and small clusters can be
already seen when considering ground state configurations
(at $T = 0$).

Before of all, increasing the number of particles leads to the
changes in the structure of lowest excitations, i.e. in the picture
of eigenvectors with smallest nonzero frequencies.
Shown in Fig.~7
are some characteristic examples of corresponding motions.
Analogous  to Coulomb clusters we found that small clusters with
{\it incommensurate}
number of particles in neighbor shells have the smallest
eigenfrequencies $\omega_{min}$.
The motion with minimal eigenfrequency $\omega_{min}$
is then corresponds to intershell rotation \cite{Bedanov}
(see Fig.~7(a),(b)).
When the number of particles is increased one can see that the
picture of a motion with the minimal eigenfrequency
evolves in such a way as to fall into a number of individual regions
of "rotations" to form "vortex/antivortex pairs" (Fig.~7(c),(d)).

An interesting effect that accompany going from small to
large clusters is the "facetting". Shells of mesoscopic
clusters have
near - circular shape, while an addition of particles leads to the
developing of well - defined facets.
This tendency is clearly seen in Fig.~8
in which the dependence of the measure of facetting $\nu_s$ on $N$
for different cluster shells $s$ is presented.
If shell number $s$ of the cluster consists of particles with
numbers $\{i_1,i_2,.. i_N\}$,
then the parameter $\nu_s$ can be defined as
\begin{eqnarray}
\nu_s = \frac{ \max\limits_{i_s} \{ |{\bf r}_{i_s}|\} }
{ \min\limits_{i_s} \{ |{\bf r}_{i_s}|\} }
\label{nu_s}
\end{eqnarray}
From Fig.~8 one can see that with increasing the number of
particles, the shape of the cluster shells approach that of
the crystal shells
defined above (appropriate values of the parameter $\nu_s$
are also shown by thick solid lines).
In support of the fact that shells
of small clusters have near - circular form, Fig.~8 shows
that minimal differences between values of parameter $\nu_s$
for crystal and cluster shells take place when the most
symmetrical crystal shell
$Cr_1$ is being filled (see $D_{17} - D_{19}, D_{32} - D_{36}$).
Instead, this difference is maximal for clusters belonging to the
other periods of "Mendeleev table"
(i.e. of $Cr_2,\; Cr_3,\; Cr_4$ basic groups).

\section{Phase transitions}
\label{transitions}

Studying thermodynamic and dynamic properties of 2D dipole clusters,
we explored the temperature region $T < 1$ and calculated the
following quantities:

{\bf 1)} Radial distribution function $g(r)$.

{\bf 2)} Pair and radial deviations:
\begin{eqnarray}
\delta_{pair} = \frac{2}{N(N-1)} \sum\limits_{i<j}
\left[ \frac{ \left< |{\bf r}_{i} - {\bf r}_{j}|^2 \right> }
{ \left< |{\bf r}_{i} - {\bf r}_{j}| \right>^2 }  - 1 \right]^{1/2}
\label{d_pair}
\end{eqnarray}

\begin{eqnarray}
\delta_{r} = \frac{1}{N} \sum\limits_{i}^{N}
\left[ \frac{ \left< |{\bf r}_i |^2 \right> }
{ \left< |{\bf r}_i | \right>^2}  - 1 \right]^{1/2},
\nonumber
\\
u_{r}^2 = \frac{1}{N} \sum\limits_{i}^{N}
\left[ \left< |{\bf r}_i |^2 \right> - \left< |{\bf r}_i |
\right>^2 \right]
\label{u_r}
\end{eqnarray}

{\bf 3)} Radial diffusion constant ${\Diff}_r$.
This quantity changes drastically
in the region of temperatures that is appropriate to the radial
disordering of the cluster and to the interchange of particles
between different shells.
For it is beyond reason to believe that a hydrodynamical
approach is valid for such small systems, we found it incorrect
to define the diffusion constant via a Fourier - image of the
velocity autocorrelation function. Instead, this quantity
was estimated from equation
\begin{eqnarray}
\left< \triangle r^2 \right> = 2 {\Diff}_r t + c
\label{D_r}
\end{eqnarray}
where $\left< \triangle r^2 \right> = \frac{1}{N} \sum_{i=1}^{N}
\left[ |{\bf r}_i(t)| - |{\bf r}_i(0)|
\right]^2
$
is a radial mean square displacement of a particle in time $t$
(in units of $\tau_0$, see above), $c$ is some constant.

{\bf 4)} Local minima distribution $\rho(\epsilon_{loc})$.
In order to estimate this histogram and to determine
the relative occupancy of different minima on the potential energy
surface, at each step of measurements several hundred steps of the
gradient minimization (see above) was come out to determine the
nearest local minimum.

In simulations of macroscopic clusters we
found it helpful to calculate the following quantities:

{\bf 5)} The number $d_n(R_c)$ of particles which are in a circle
of radius $R_c$ and have $n$ nearest neighbors.
The Voronoi construction was applied
to find nearest neighbors of each particle.

{\bf 6)} Static structure factor $S({\bf k})$
\begin{eqnarray}
S({\bf k}) = \frac{1}{N} <\rho_{{\bf k}} \rho_{-{\bf k}}>,
\nonumber
\\
\rho_{{\bf k}} = \sum\limits_{i=1}^{N}
\exp{ (\jmath {\bf k} {\bf r}_i )},\;
\jmath^2 = -1
\label{S_k}
\end{eqnarray}

A usual way of studying the intershell orientational disordering
in clusters is an analysis of relative angle deviations
of shells, in analogy with quantities (\ref{d_pair}) - (\ref{u_r}).
In this approach the temperature $T_{s_1 s_2}$ of orientational
"melting" of shells $s_1$ and $s_2$ of a cluster is defined as that
at which there is a sharp increase in the value of
appropriate relative angle deviations. We follow a
different method of finding the disordering temperature
$T_{s_1 s_2}$ as that at which the
{\it mutual orientational (order) parameter} of shells $s_1$ and
$s_2$ vanishes. We define this quantity as follows:
for each shell $s$ of $N_s$ particles we consider complex -
valued quantity $\psi_s$:
\begin{eqnarray}
\psi_s =
\frac{1}{N_s} \sum\limits_{i_s} \exp{(\jmath N_s \varphi_i)}
\label{psi}
\end{eqnarray}
The sum in (\ref{psi}) is extended over all particles possessed
by shell $s$. The {\it mutual orientational (order) parameter}
is then defined as
\begin{eqnarray}
g_{s_1 s_2} = < \psi_{s_1} \psi_{s_2}^*>
\label{g_s1s2}
\end{eqnarray}
It is obvious, that $g_{s_1 s_2}$ disappeares at the point of
relative disordering (slipping) of shells $s_1$ and $s_2$
(in the vicinity of this the very point there will be also a sharp
increase in relative angle deviations).
The quantity $g_{s s} = <|\psi_s|^2>$ will be a
measure of the intrashell order.
We note that quantities $\psi_s$ and $g_{s_1 s_2}$
are analogous to orientational parameter $\psi_6$ and
correlation function $g_6(r)$ in infinite 2D system, where
vanishing (when the translational order is absent) of the
correlation function, $g_6(r) \to 0$ as $r \to \infty$,
indicates the relative orientational disordering of distant parts
of a system.

\subsection{Mesoscopic clusters}
\label{small}

The distinctive property of mesoscopic clusters is the presence
of {\it two types} of disordering effects in these systems:
\cite{Mandelstam}$^{,}$\cite{Rakoch_log,Rakoch_dip}
an intershell orientational disordering
(an orientational melting of shells $s_1$ and $s_2$ at temperature
$T_{s_1 s_2}$)
and a radial disordering (a total melting at temperature $T_f$
that is larger than any one of orientational melting temperatures).
An analysis of eigenfrequencies and eigenvectors for the normal
modes of small clusters shows that the motions with small lowest
nonzero eigenfrequencies $\omega_{min}$ correspond to intershell
rotation (see Fig.~7).
Such clusters will have small temperatures $T_{s_1 s_2}$
of intershell disordering, at which shells start to rotate relative
to each other losing their mutual orientational order.
Note that, in contrast to the case of large clusters, an intershell
melting in small clusters takes place for {\it all} pairs of shells,
i.e. there exist "melting" temperatures $T_{2 1}$, $T_{3 2}$,
$T_{4 3}$... .
In clusters of large number of particles ($N > 37$) an orientational
disordering of only the outer shell is possible. An analogous
observation was made for Coulomb clusters. \cite{Bedanov}

Shown in Fig.~9(b) are dependencies of the values of mutual
orientational parameters $g_{2 1}$ and $g_{3 2}$ {\it vs.}
temperature for three - shell cluster $D_{24}$.
Pair and radial deviations
$\delta_{pair}(T)$ and $u^2_r(T)$
(see (\ref{d_pair}),(\ref{u_r})) are also plotted.
This figure shows that for cluster $D_{24}$ one can define two
temperatures $T_{3 2} = (5 \pm 0.5) \cdot 10^{-4}$ and
$T_{2 1} = (3 \pm 0.05) \cdot 10^{-4}$
that are correspond to orientational melting of shells
$\{3,2\}$ and $\{2,1\}$.

Fig.~9(b) (see also Fig.~10) shows that pair and radial deviations
(\ref{d_pair}), (\ref{u_r}) are also sensitive to intershell
rotation.
One can see the regions of sharp increases in the values of these
quantities that coincide with the regions of vanishing of mutual
orientational order.
The sensitivity of radial and pair deviations to the
orientational disordering is due to the "breathing"
of the cluster shells on their rotation.
This breathing can be clearly seen if one trace the motion of a
system along a "reaction path", \cite{Quapp}
the most probable (of the weakest ascent in $2N$ - dimensional
space) trajectory which is appropriate to the intershell rotation.

Temperature $T_{3 2} \approx 0.05 \pm 0.003$ of the orientational
"melting" of the third shell of four - shell cluster
$D_{35}(\underline{1,6,12},16)$
is only slightly lower than the temperature $T_f = 0.065 \pm 0.005$
of the total melting at which
the radial order disappears (see Fig.~10).
It is obvious
that the temperature of an orientational intershell disordering
will to a large extent depend on the distribution of particles
throughout shells.
Particularly, when the particle numbers
in neighbor shells $s_1$ and $s_2$ are {\it commensurate},
temperature $T_{s_1 s_2}$ of an orientational "melting" should be
maximal (irrespective of whether these shells are inside the
cluster or a pair of external shells is considered).
Calculations do show that an addition of as little as one
particle to small cluster
$D_{19}(\underline{1,6,12})$ with completely filled
crystal shells of
$Cr_1$ type leads to abrupt decreasing of temperature $T_{3 2}$
of orientational "melting".
Indeed, temperature
$T_{3 2} \approx 0.03 \pm 0.002$ of $D_{19}$ cluster is
very nearly equal to that of the total melting
$T_f \approx 0.038 \pm 0.003$ (at which particles begin to
interchange between shells).
System $D_{20}(1,7,12)$ becomes orientationally disordered
at much smaller temperatures $T_{3 2} < 5 \cdot 10^{-6}$.

\subsection{Macroscopic clusters}
\label{large}

The most interesting when considering macroscopic clusters
(at $N > 37$) is the question about the manner in which their
melting temperatures $T_f$ approach the temperature
$T^{inf}_c = k_b T / (d^2 n^{3/2}) = 0.089 \pm 0.002$
of first order phase transition in 2D
infinite dipole system
(here $n$ stands for the density of particles in the system).
\cite{2DPhonon,2DDip}
But this analysis is made difficult
by an uncertainty in the concept of the "density" of a cluster.
This uncertainty does not enable us to use the natural for infinite
systems unit of length $a = 1/\sqrt{n}$. When converting from
"cluster" to "infinite system" temperature, one can introduce a
{\it mean} density of a system as $n = N / \pi R^2$
where $R$ is the radius of the cluster.
The uncertainty in $R$ leads to the ambiguity
of such definition of the cluster density.
Moreover, if we deal with a cluster of a large enough number of
particles the density will considerably vary with the distance
to the center of the system.
We will discuss this question in some more detail below.

To define an effective (mean) density of a cluster of not
very large number of particles ($N < 50$), when the
nonuniformity of the density pointed out above is not essential,
we made use of calculations of the static structure factor
(\ref{S_k}).
Let us consider the distance $|\delta {\bf k}|$ to the maximum
of the structure factor, i.e. to the first Bragg peak for
a "solid" cluster (when $T<T_f$) or to the Lorentz peak in
the case of a "liquid" cluster (when $T>T_f$).
The characteristic distance between
particles in a cluster can be defined then as
$a = 2\pi / |\delta {\bf k}| =
\left[ 2 / (n \sqrt{3}) \right]^{1/2}$.
In such a manner the corresponding temperature of 2D infinite
system can be estimated as
\begin{eqnarray}
T^{inf}  = \frac{ k_b T }{ D^2 n^{3/2} } \approx
T \frac{1}{0.8} \left[ \frac{|\delta {\bf k}|}{2 \pi} \right]^{3}
\label{T_inf}
\end{eqnarray}
It turned out that for systems $D_{36} - D_{52}$
the estimation $T \approx T^{inf}$ works with a reasonable accuracy.

One of the most representative quantities for 2D
infinite dipole system is excess energy $\epsilon$ that
exhibits a jump of $\triangle \epsilon \approx 0.04$ at
a temperature of phase transition $T^{inf}_c$. \cite{2DDip}
Our calculations show (see Fig.~11) that such a sharp increase
in an excess energy {\it does not take place} at least
for clusters with $N < 50$.

Yet another quantity that is commonly used in analyzing phase
transitions in infinite systems is structure factor $S({\bf k})$,
Eq.~(\ref{S_k}).
The lattice to liquid transition is identified through the vanishing
of the first Bragg peak $S({\bf k}),\; {\bf k} \approx {\bf q}_1$.
We found that the magnitude of this peak ( the maximum of the
structure factor in the region $|{\bf k}| > k_c \sim \pi$)
is acutely sensitive to
the disordering in clusters.
The peak at small wave vectors
$|{\bf k}| < \pi$ that exists at any temperature due to
finite size of the system involved is of no interest.
Some characteristic examples of the behavior of the structure
factor as a function of temperature are given in Fig.~12.

An independent quantity a sharp change in which testifies about
the order/disorder transition is the radial diffusion constant
${\Diff}_r$, see equation~(\ref{D_r}).
Fig.~13 presents the results of calculations of this quantity,
performed for clusters $D_{39}$, $D_{40}$ and $D_{45}$.
Each data point in this figure was obtained by averaging over
results of 5 independent experiments in which the evolution of
the system was observed in a time that is typical for the
diffusion of a particle for a distance of $\triangle r \sim 1$.
The result of one of such experiments is shown in the insert of
the figure.

An analysis of temperature dependencies of above quantities
(\ref{d_pair}) - (\ref{S_k}) enables us to plot the
total melting temperature $T_f(N)$ as a function of the number
of particles in the cluster.
This curve is shown in the insert of Fig.~14.
Also shown are temperatures $T^{inf}_f(N)$ of corresponding
infinite system that are estimated with the help of Formula
(\ref{T_inf}). Fig.~14 enables us to point out some
characteristic features of the $N$ dependence of the
disordering temperature.
{\bf i)}~Melting of all clusters studied takes place at
temperatures much more smaller than temperature
$T_c^{inf} \approx 0.1$
of the phase transition in an infinite system.
{\bf ii)}~The most interesting peculiarity of function
$T_f(N)$ is that it is nonmonotonic.
Most likely, this peculiarity is primarily connected
with the fact that
a cluster of $N$ particles can be considered as a part of a
{\it strongly deformed} crystal lattice. Hence one studies
the melting of a part of imperfect crystal with "frozen-in"
interstitials and dislocations (see Fig.~2(a)), the melting
temperature
being a function of the number of such defects (of the value of
an initial deformation of a cluster as a part of crystal).
Obviously, the number of defects in the ground state
configuration will be minimal in the case of "magic" cluster
(see above) with a maximal number of filled crystal shells.
An analysis of the data presented in Table~\Roman{Table1}
and in Fig.~14 shows that (at least for $N < 52$) the "magic"
clusters do have anomalously high melting temperatures.

Of course, the melting temperature is not a strictly defined
quantity for finite systems.
It makes sense
to consider an interval $\triangle T(N)$ of temperatures in which
the disordering takes place. For clusters of large number of
particles that repel each other and are in a confinement potential
there is another, more important, reason that does not enable us to
introduce a concrete melting temperature of a phase transition
(of a total disordering).
The point is that large clusters (at $N > 50$) are irregular in both
the density of particles and the density of defects.
These
densities can vary appreciably with the distance $r$ to the center
of the system \cite{Note1} to lead to a strong $r$ dependence
of the melting temperature.

Let us consider this dependencies in some more detail.
Results of numerical simulations show that large clusters
can be divided into the following regions: \cite{Koulakov}
{\bf (a)}~the "core" of a cluster, a region adjacent to the
center of the island and filled with a number of completed
crystal shells;
{\bf (b)}~the "layer of defects", the ring in which
dislocations and disclinations are situated and which is
"concentric" to the surface of the cluster;
{\bf (c)}~the region between the surface of the island and
the layer of defects.
One can assume that temperatures of disordering in the
regions considered above will differ greatly.
Our calculations approve this assumption.
As an example, we calculated the number of disclinations $d_n(r)$
and the radial diffusion constant ${\Diff}_r(r)$
as functions of the distance $r$ to the center of the cluster
$D_{80}$.
It turned out that temperature $T_f^{inf}(r < r_c)$
of the disordering of a "core", the part of the cluster
belonging to the circle of the radius $r_c \approx 2.5$,
was independent of $r$:
$T_f^{inf}(r < 2.5) \approx 0.07 \pm 0.01$.
This temperature is slightly below $T_c^{inf}$.
One may suppose that as $N$ increases, the scaled temperature
of core disordering $T_f^{inf}(N; r<r_c)$ approaches the
temperature of phase transition in 2D infinite dipole system:
$T_f^{inf}(N; r<r_c) \to T_c^{inf}, \quad N \to \infty$.

The melting in other two regions of the cluster appeared to take
place at much less temperature:
$T_f^{inf}(r > r_c) = 0.01 \pm 0.005$.
Obviously, the temperature of disordering in regions {\bf (b)} and
{\bf (c)} will strongly depend on the number
of disclinations and dislocations in the layer of defects
(in the region {\bf (b)}) as well as on the strength
of initial deformations in the region {\bf (c)}.
Thus we argue that: \\
{\bf i)} defined as a point of drastic changes in pair and radial
deviations, diffusion constant and of
disappearance of the first Bragg peak,
neither "melting" temperature $T_f(N)$ nor scaled to 2D infinite
system melting temperature $T_f^{inf}(N)$ tends to
first - order phase transition
temperature in infinite 2D system as the number of particles in the
cluster increases. \\
{\bf ii)} The function $T_f^{inf}(N)$ (as well as $T_f(N)$) is not
monotonic.
The magic number clusters, having minimal number of defects,
have maximal melting temperatures. \\
{\bf iii)} One can introduce the temperature
$T_f^{inf}(N; r<r_c) \to T_c^{inf}, \quad N \to \infty$ which is
the (scaled to 2D infinite system) melting temperature of a core,
the free of defects central region of a cluster. \\
These qualitative predictions are presented in Fig.~14.

\subsection{Cluster $D_{37}$}
\label{D_37}

As have been pointed out above, system $D_{37}$
can be considered as an intermediate one between mesoscopic and
macroscopic clusters.
An analysis of the global minimum
configuration (see Fig.~5) and of the structure of normal motion
with the minimal eigenfrequency $\omega_{min}$ leads to the
seemingly unambiguous conclusion about the absence of the
intershell orientational disordering.
However, our calculation shows that the melting in this system
is not one - step process;
instead, the orientational melting takes
place at the temperature $T=T^* \approx 0.01$ that is
approximately three times lower than that of total melting
$T_f \approx 0.03 \pm 0.005$.
The orientational disordering of {\it all} pairs of shells takes
place at temperature $T^*$.

The reason of this "anomalous" behavior of cluster $D_{37}$
can be elucidated by considering the structure of
the lowest local minima configurations (i.e. the configurations
with the lowest energy).
We have noted above (see Fig.~5(c)) that global and local
minima configurations differ in a {\it symmetry}, namely,
any breaking of the ground state configuration down into shells
leads to the necessity of viewing one or two
particles (see Chapter~\ref{ground_config}) as interstitials
interposed between shells of the cluster.
Only two first shells
are well defined: $D_{37}(\underline{1,6},...)$.
On the contrary, the configuration of the first excited state
(of the lowest local minima with energy
$\epsilon_{loc}^{(1)} \approx 8.3325$)
is rather symmetrical and has the clear shell structure:
$D_{37}^{(1)}(1,7,13,16)$.
As shells of the cluster in this local minimum
are not facetted, it is not surprising that the motion
(in this local minimum) with the lowest
eigenfrequency $\omega_{min}^{(1)} \approx 0.4$ corresponds to the
intershell rotation.
This is illustrated in Fig.~5(c).

Plotted in Fig.~15 is the probability of the central particle
having six neighbors. This quantity can be defined as
$d_6(0.5)$, see above. Fig.~15 shows
that at temperatures $T > T^{*} = 0.01 \pm 0.002$
there is a finite probability to find a system in the vicinity
of the local minimum $D_{37}^{(1)}(1,7,...)$.
Yet another illustration of
this rearrangement is the change in histogram
$\rho(\epsilon_{loc})$ of the local minima distribution shown
in the insert of Fig.~15.
One can see that at $T = T^{*}$
a pike at energy $\epsilon_{loc}^{(1)}$ appears.
Further increasing the temperature leads to the
occupancy of others local minima and, at $T > 0.3$,
the radial disordering takes place,
particles interchange between shells and the radial
distribution function is washed out (see Fig.~15,16).
So, at $T > T^{*}$, a part of time the system lives in
the neighborhood of
the "symmetrical" local minimum that is {\it unstable} against
intershell rotation at such high temperatures.
This intershell disordering
also manifests itself as a sharp increase in relative angular
intershell deviations at this the very temperature $T = 0.01$.

From results of studies of atomic clusters \cite{Berry,Chakravarty}
it is known that clusters of some specific number of particles
(so called "magic" $LJ$ clusters such as $LJ_{13},\; LJ_{19}$, see
work [\onlinecite{Berry_1}] and references therein) have distinct
regions of coexistence of "liquidlike" and "solidlike" forms.
In a certain
temperature interval there is nonzero probability to find
the cluster either in a "liquidlike" or in a "solidlike" form.
It was shown that this unusual property owes its origin
to features of the multidimensional potential energy surface
and stems from an existence of regions
(well - separated by high barriers) in phase space
with distinct physical properties and appreciable residence time
for the system in each region.

A picture of disordering in cluster $D_{37}$
with increasing temperature  closely resembles the picture
of structural transitions in magic number atomic clusters.
Of course, at sufficiently high temperatures,
at $T > T^* = 0.01$, the system can be found in one of two
distinct regions in phase space with a barrier between them.
This equilibrium can be described as:
$$
D_{37}(\underline{1,6},\tilde{1},13,16) \stackrel{K}{\rightleftharpoons}
D_{37}(1,7,13,16),
$$
$$
K = \frac{ \left[ D_{37}(1,7,13,16) \right] }{
\left[ D_{37}(\underline{1,6},\tilde{1},13,16) \right] } =
e^{ - \triangle F(T) / T  }
$$
where $\triangle F(T)$ is free energy difference of "symmetrical"
and "fully ordered" states of the cluster.
At $T < 0.01$ constant of equilibrium $K \sim 1 - d_6(0.5)$
is equal to zero and increases markedly as the temperature is
increased (see Fig.~15).

The analysis above shows that,
with the cluster $D_{37}$ being radial ordered, the region
$0.01 < T < 0.03$ {\it can not}
be considered as the region of coexistence of its "solidlike"
and "liquidlike" forms.
Instead, this temperature interval
can be though of as the region of coexistence of
"solidlike" and "orientationally disordered" forms.
It is the presence of such region that enables us to consider
the system $D_{37}$ as an intermediate one that has
features of both mesoscopic and macroscopic clusters.

To conclude of this paragraph we note that
the number of $D_{37}$ - like "anomalous" systems can
be varied by tuning the characteristic
range of an interaction potential.
Increasing the temperature in such systems can lead to
structure rearrangements (with changes in symmetry),
changes in the shell distribution or in types of defects.
Consequently, above consideration can be helpful in the study
of thermodynamic properties of the clusters,
parameters of interparticle potential of which can be varied in wide
limits, e.g. of dusty plasma clusters, \cite{Dp,Fortov}
of the system of electrons in a semiconductor dot with
near metal electrodes, \cite{West,Koulakov} etc.

\section{Conclusion}
\label{conclusions}

We have presented the results of a numerical simulation of
finite 2D dipole system in the parabolic confinement.
Ground state configurations
and the spectrum of normal modes of clusters of $N \le 80$
particles have been found.
An addition of particles to the system
leads to a gradual filling of crystal shells of different symmetry.
The clusters with minimal energies and maximal lowest
eigenfrequencies ("magic" clusters) are that with the maximal number
of completely filled crystal shells and (in the case of large clusters,
$N > 40$) with equal numbers of particles on the last two shells.
Such clusters have also well developed facets.

The character of disordering with increasing the temperature
is markedly different for mesoscopic ($N < 37$) and macroscopic
($N > 37$) clusters.
Particularly, mesoscopic clusters are characterized by
an existence of two types of disordering effects:
orientational and full (radial).
Depending on the
degree to which crystal shells are completed the temperatures of
orientational "melting" of different pairs of shells can differ
greatly from each other.
Orientational melting in large systems is absent.

An analysis of the local minima distribution of the system
$D_{37}$ as a function of temperature have shown
that this system has features of both mesoscopic and
macroscopic clusters.
Namely, there is a region of temperatures in which
cluster is in a dynamical equilibrium between "solidlike"
and radial ordered but orientationally disordered forms.

The system of large number of particles is nonuniform:
both the characteristic interparticle distance and the density
of defects are functions of a distance $r$ from the
center of the system.
As a consequence, the temperature at which
a sharp increase in radial and pair deviations, vanishing of the
first Bragg peak take place and the diffusion of particles appear
does not approach temperature $T_c^{inf}$
of first - order phase transition
in 2D infinite dipole system as the number of particles is
increased.
Scaled to infinite system, "melting" temperature $T^{inf}_f(N)$
is a nonmonotonous function of the number of particles
and is maximal for magic number clusters.
By taking into account the $r$ dependence of the interparticle
distance, the melting of the "core" of the cluster
(i.e. its free of defects central region $r < r_c(N)$)
can be mapped to the melting of 2D infinite system
with appropriate scaled temperature
$T^{inf}(N; r < r_c) \to T^{inf}_c$ as $N$ increases.

\vspace{1cm}

{\bf Acknowledgments}

We wish to thank A.M.~Popov and S.A.~Verzakov for fruitful
discussion. The work has been supported by Russian Foundation of
Basic Research, INTAS and the Program "Physics of
Solid Nanostructures".

\onecolumn

\newpage

{
Table~\Roman{Table1}. \\
Ground state configurations of 2D dipole clusters $D_N$
in a harmonic confining. Shown are shell configurations $\{N_1,N_2,...\}$,
types of crystal shells (see in Text) and excess energy
$\epsilon = E / N$. \\
}
\renewcommand{\baselinestretch}{1.4}
\begin{footnotesize}

\begin{tabular}{p{1cm}p{2.2cm}p{1.2cm}p{3cm}|p{1cm}p{2.4cm}p{1.2cm}p{2cm}}
\hline
\hline
$N$ & $\{N_1,N_2,...\}$ &  & $\epsilon$  & $N$ & $\{N_1,N_2,...\}$ &  & $\epsilon$ \\
\hline
1 & 1 &             &                  $0$        & 41 & \underline{3,9},14,15 & $\;\;Cr_3$ &       $8.93397 $ \\
2 & 2 &             &                  $0.64660 $ & 42 & \underline{3,9},14,16 & $\;\;Cr_3$ &       $9.08148 $ \\
3 & 3 &             &                  $1.01394 $ & 43 & \underline{3,9,15},16 & $\;\;Cr_3$ &       $9.22705 $ \\
4 & 4 &             &                  $1.38021 $ & 44 & \underline{3,9,15},17 & $\;\;Cr_3$ &       $9.37309 $ \\
5 & 5 &             &                  $1.75713 $ & 45 & \underline{4,10},15,16 & $\;\;\;Cr_4$ &      $9.51691 $ \\
6 & 1,5 &  &                           $2.04829 $ & 46 & \underline{4,10},15,17 & $\;\;\;Cr_4$ &      $9.66006 $ \\
7 & \underline{1,6} & $Cr_1$  &       $2.32591 $ & 47 & \underline{4,10,16},17 & $\;\;\;Cr_4$ &      $9.80303 $ \\
8 & 1,7 & $Cr_1$  &                   $2.63542 $ & 48 & \underline{4,10,16},18 & $\;\;\;Cr_4$ &      $9.94353 $ \\
9 & 2,7 & $\;Cr_2$  &                   $2.92373 $ & 49 & 1,5,11,16,16 &           $Cr_1$ &      $10.08318$ \\
10 & 3,7 & $\;\;Cr_3$  &                  $3.19012 $ & 50 & \underline{1,6},11,16,16 & $Cr_1$ &    $10.21803$ \\
11 & 3,8 & $\;\;Cr_3$  &                  $3.41972 $ & 51 & \underline{1,6,12},16,16 & $Cr_1$ &    $10.35593$ \\
12 & \underline{3,9} & $\;\;Cr_3$  &      $3.66665 $ & 52 & \underline{1,6},11,17,17 & $Cr_1$ &    $10.49035$ \\
13 & 4,9 & $\;\;\;Cr_4$  &                  $3.89493 $ & 53 & \underline{1,6,12},17,17 & $Cr_1$ &    $10.62096$ \\
14 & \underline{4,10} & $\;\;\;Cr_4$ &      $4.13543 $ & 54 & \underline{1,6,12},17,18 & $Cr_1$ &    $10.75525$ \\
15 & 5,10 &         &                  $4.35999 $ & 55 & \underline{1,6,12,18},18 & $Cr_1$ &    $10.88617$ \\
16 & 1,5,10 &  &                       $4.56558 $ & 56 & \underline{1,6,12,18},19 & $Cr_1$ &    $11.02094$ \\
17 & \underline{1,6},10 & $Cr_1$ &    $4.77272 $ & 57 & \underline{1,6,12,18},20 & $Cr_1$ &    $11.15521$ \\
18 & \underline{1,6},11 & $Cr_1$ &    $4.97257 $ & 58 & \underline{1,6,12,18},21 & $Cr_1$ &    $11.28939$ \\
19 & \underline{1,6,12} & $Cr_1$ &    $5.18009 $ & 59 & \underline{2,8},13,18,18 & $\;Cr_2$ &    $11.41878$ \\
20 & 1,7,12 & $Cr_1$ &                $5.38833 $ & 60 & 3,8,13,18,18 & $\;\;Cr_3$ &                $11.54741$ \\
21 & 2,7,12 & $\;Cr_2$ &                $5.59048 $ & 61 & \underline{2,8,14},18,19 & $\;Cr_2$ &    $11.67875$ \\
22 & \underline{2,8},12 & $\;Cr_2$ &    $5.77969 $ & 62 & \underline{2,8,14},19,19 & $\;Cr_2$ &    $11.80329$ \\
23 & 3,8,12 & $\;\;Cr_3$ &                $5.96866 $ & 63 & 3,8,14,19,19 & $\;\;Cr_3$ &                $11.92863$ \\
24 & 3,8,13 & $\;\;Cr_3$ &                $6.14713 $ & 64 & \underline{3,9},14,19,19 & $\;\;Cr_3$ &    $12.05160$ \\
25 & \underline{3,9},13 & $\;\;Cr_3$ &    $6.32561 $ & 65 & \underline{3,9,15},19,19 & $\;\;Cr_3$ &    $12.17598$ \\
26 & 4,9,13 & $\;\;\;Cr_4$ &                $6.50834 $ & 66 & \underline{3,9},14,20,20 & $\;\;Cr_3$ &    $12.30108$ \\
27 & 4,9,14 & $\;\;\;Cr_4$ &                $6.68410 $ & 67 & \underline{3,9,15},20,20 & $\;\;Cr_3$ &    $12.42251$ \\
28 & \underline{4,10},14 & $\;\;\;Cr_4$ &   $6.85654 $ & 68 & \underline{3,9,15},20,21 & $\;\;Cr_3$ &    $12.54674$ \\
29 & 5,10,14 & &                       $7.03598 $ & 69 & \underline{4,10},15,20,20 & $\;\;\;Cr_4$ &   $12.66865$ \\
30 & 5,10,15 &  &                      $7.20543 $ & 70 & 5,10,15,20,20 &  &                      $12.78866$ \\
31 & 1,5,10,15 & $Cr_1$ &             $7.36745 $ & 71 & 1,5,10,15,20,21 &  &                    $12.90960$ \\
32 & \underline{1,6,12},13 & $Cr_1$ & $7.52917 $ & 72 & \underline{4,10,16},21,21 & $\;\;\;Cr_4$ &   $13.03147$ \\
33 & \underline{1,6,12},14 & $Cr_1$ & $7.68741 $ & 73 & 1,5,11,16,20,20 &  &                    $13.15305$ \\
34 & \underline{1,6,12},15 & $Cr_1$ & $7.84408 $ & 74 & 1,5,11,16,21,20 &  &                    $13.27021$ \\
35 & \underline{1,6,12},16 & $Cr_1$ & $8.00361 $ & 75 & 1,5,11,16,21,21 &  &                    $13.38454$ \\
36 & \underline{1,6,12},17 & $Cr_1$ & $8.16885 $ & 76 & \underline{1,6},11,16,21,21 & $Cr_1$ & $13.50135$ \\
37 & \underline{1,6},$\tilde{1}$,13,16 & $Cr_1$ &             $8.33111 $ & 77 & \underline{1,6,12,18},20,20 & $Cr_1$ & $13.61939$ \\
38 & \underline{2,8},13,15 & $\;Cr_2$ & $8.48550 $ & 78 & \underline{1,6,12},17,21,21 & $Cr_1$ & $13.73279$ \\
39 & 3,8,13,15 & $\;\;Cr_3$ &             $8.63869 $ & 79 & \underline{1,6,12,18},21,21 & $Cr_1$ & $13.84553$ \\
40 & \underline{3,9},14,14 & $\;\;Cr_3$ & $8.78617 $ & 80 & \underline{1,6,12},17,22,22 & $Cr_1$ & $13.96137$ \\
\end{tabular}
\end{footnotesize}
\\

\twocolumn

Fig.~1 \\
Different parts of 2D hexagonal lattice produce the basis for
ground state configurations of the majority of dipole clusters
(see Table~\Roman{Table1}).
Shown are examples of such basic groups of crystal shells.
Four most symmetrical groups of three shells each
is presented.

\vspace{0.7cm}

Fig.~2 \\
Ground state configurations of dipole clusters:
a) $D_{44}$; b) $D_{47}$; c) $D_{52}$; d) $D_{59}$.
Squares denotes particles with 5 nearest neighbors.

\vspace{0.7cm}

Fig.~3 \\
Two possible breakups of cluster $D_{62}$ into shells:
a) in such a way as to maximize the number of completed
$Cr_4$ shells.\cite{Koulakov}
b) so that the outer shell be well - defined and closed.

\vspace{0.7cm}

Fig.~4 \\
The second and the first (in the insert of the Figure)
"derivatives" of specific energy $\epsilon(N)$
with respect to $N$.

\vspace{0.7cm}

Fig.~5 \\
Cluster $D_{37}$. a) The ground state configuration
$D_{37}(\underline{1,6},\tilde{1},13,16)$.
b) the eigenvector of the cluster motion in the global minimum
with the minimal nonzero eigenfrequency $\omega_{min} \approx 0.58$
c) the picture of the motion in the lowest local minimum
$D_{37}^{(1)}(1,7,13,16)$ with a minimal eigenfrequency
$\omega_{min}^{(1)} \approx 0.4$. \\
Five - coordinated particles are marked by squares.

\vspace{0.7cm}

Fig.~6 \\
Minimal nonzero eigenfrequencies $\omega_{min}$ for the normal modes
of $N$ - particle dipole clusters.

\vspace{0.7cm}

Fig.~7 \\
Characteristic examples of normal motions with eigenfrequencies
$\omega_{min}$ (see Fig.~6).
a) $D_{10}$; b) $D_{35}$; c) $D_{55}$; d) $D_{56}$.

\vspace{0.7cm}

Fig.~8 \\
The measure of faceting $\nu_s(N)$ is shown for different
shells $s$. Appropriate values for crystal
shells of different types are plotted by thick solid lines.

\vspace{0.7cm}

Fig.~9 \\
Three - shell cluster $D_{24}(3,8,13)$.
The results of calculations of the mutual otientational
parameter (\ref{g_s1s2}) and of deviations (\ref{d_pair}),(\ref{u_r})
as functions of temperature $T$ are presented. \\
The data are connected to guide the eyes. If not present, error bars
are smaller than the size of the data point.

\vspace{0.7cm}

Fig.~10 \\
Four - shell cluster $D_{35}(\underline{1,6,12},16)$.
Mutual orientational parameter $Re |g_{3 2}|$
and relative deviations $\delta_{pair}(T)$ vs. temperature $T$.

\vspace{0.7cm}

Fig.~11 \\
Temperature dependence of excess energy $\epsilon(T)$
for a number of dipole clusters $D_{N}$.
In the insert the excess energy of $D_{37}$ cluster in
the temperature region $T \in [0.02, 0.02]$ is shown.

\vspace{0.7cm}

Fig.~12 \\
The magnitude of the first Bragg peak as a function of $T$
for clusters $D_{40}$ and $D_{45}$. Shown also is 2D topography of
the structure factor $S({\bf k}), \quad S({\bf k})<0.5$
at different temperatures. "Solidlike" to
"liquidlike" transition is accomplished by the washing out of
$S({\bf k})$ in regions ${\bf k} \approx {\bf q}_1$
of reciprocal lattice vectors.

\vspace{0.7cm}

Fig.~13 \\
The radial diffusion constant (\ref{D_r}) vs. temperature $T$.
The insert shows a typical diffusive motion of particles when
the cluster is in disordered state ($T \ge T_f$). Results of
the linear fit (see Equation (\ref{D_r})) are also given.

\vspace{0.7cm}

Fig.~14 \\
The total melting temperatures $T_f(N)$ and scaled from "cluster" to
"infinite system" temperatures $T^{inf}_f(N)$ (\ref{T_inf})
as functions of number of particles $N$.
Points correspond to the results of present simulations.
The dotted line depicts a possible behavior of $T^{inf}_f(N)$
at $N > 50$.
Temperature of first - order phase transition in an infinite
2D dipole system $T_c^{inf}$ is shown with the help of a dashed line.
Temperature $T_f^{inf}(N; r<r_c)$ of the disordering of a core
is assumed to tend to $T_c^{inf}$ as $N$ increases.

\vspace{0.7cm}

Fig.~15 \\
The probability that the central particle of cluster $D_{37}$
has six neighbors as a function of temperature $T$.
In the insert local minima distributions for different temperatures
are shown.
One can see that at $T > 0.01$ a group of local minima with energies near
$\epsilon^{(1)}$, the energy of the lowest local minimum, are being occuped.

\vspace{0.7cm}

Fig.~16 \\
Radial distribution function $g(r)$ of cluster $D_{37}$.

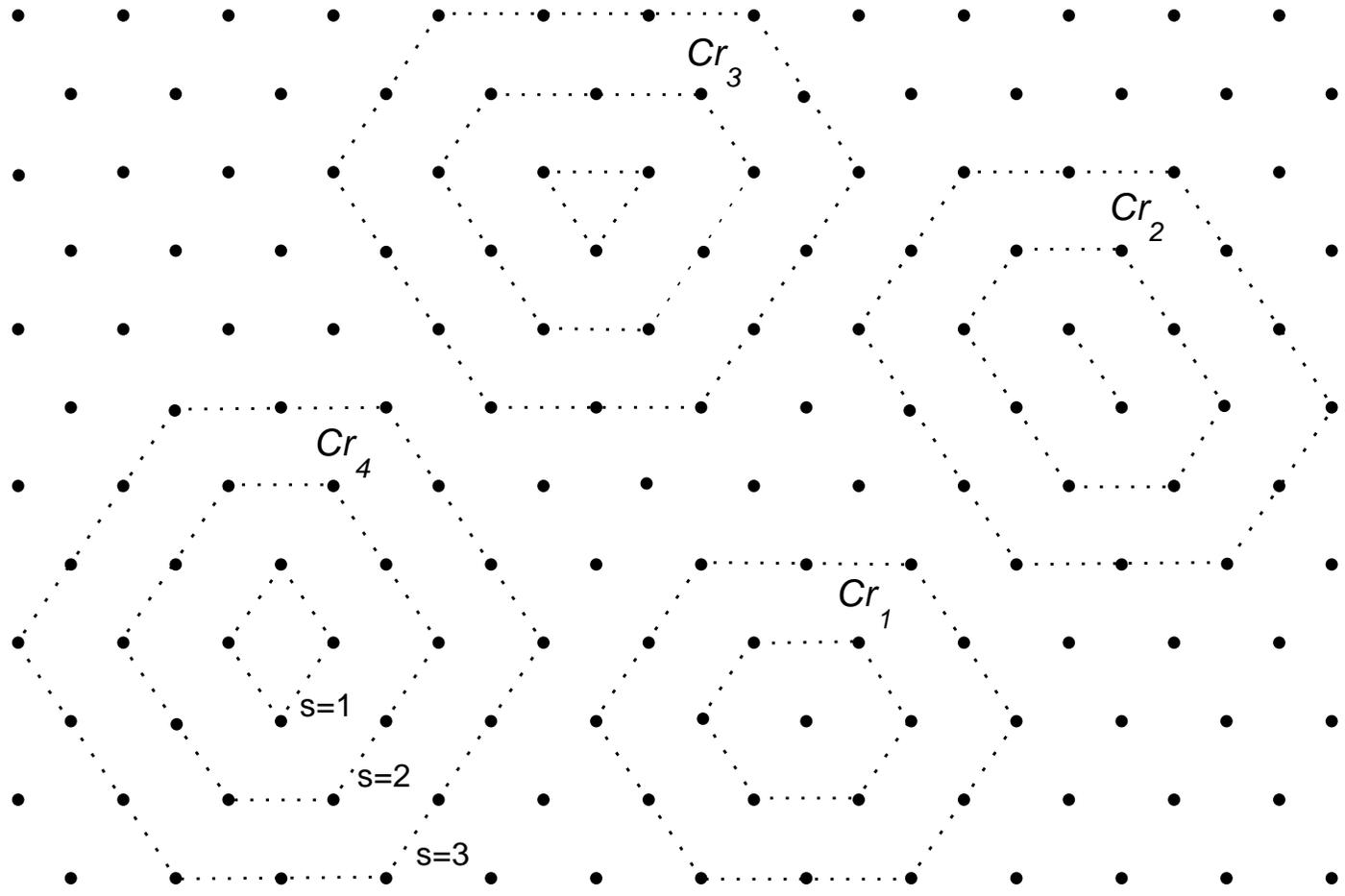
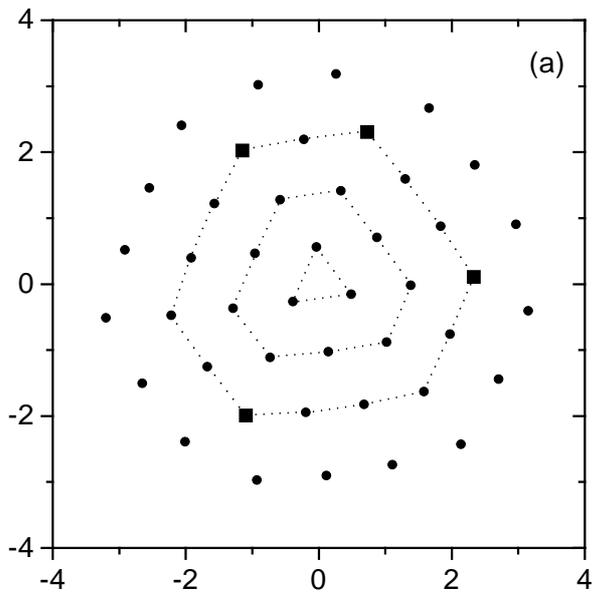
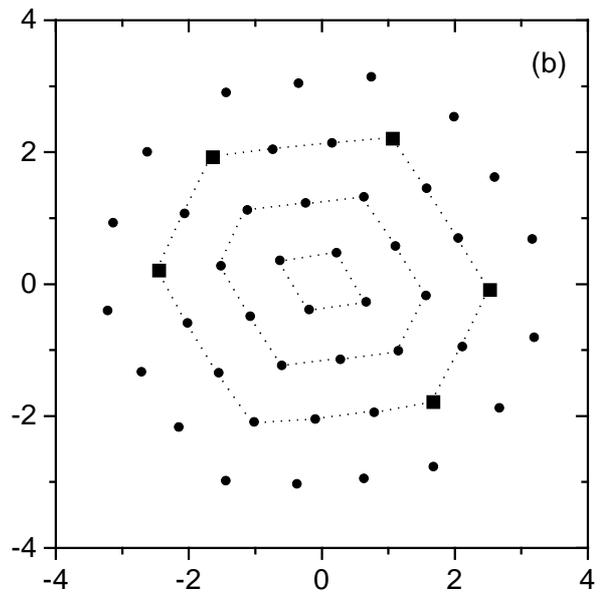
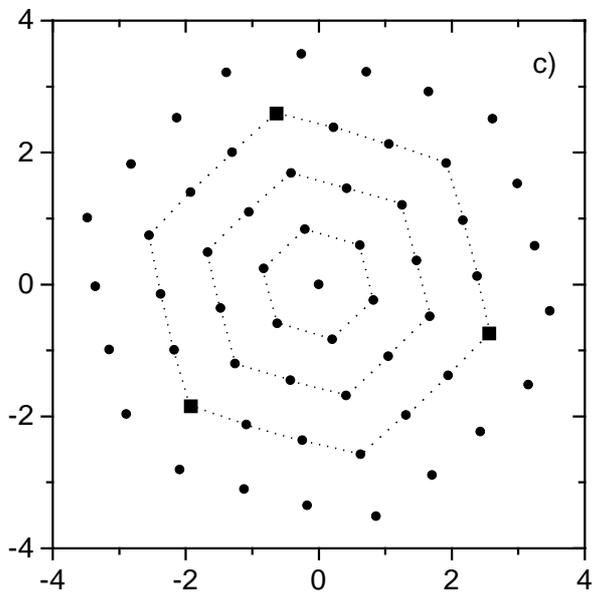
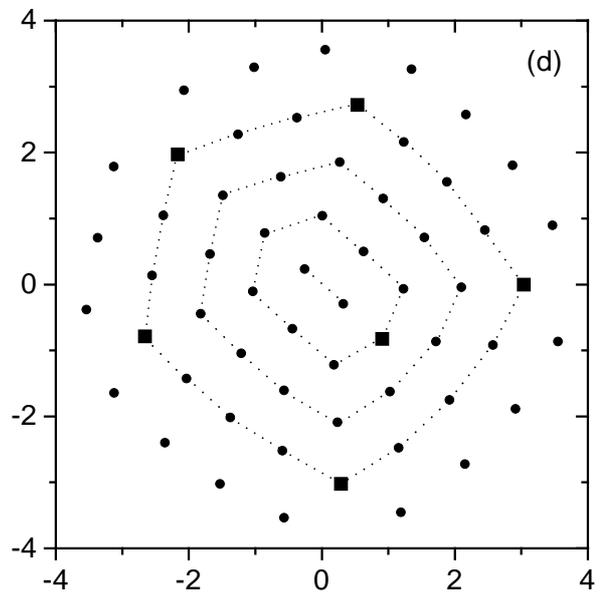
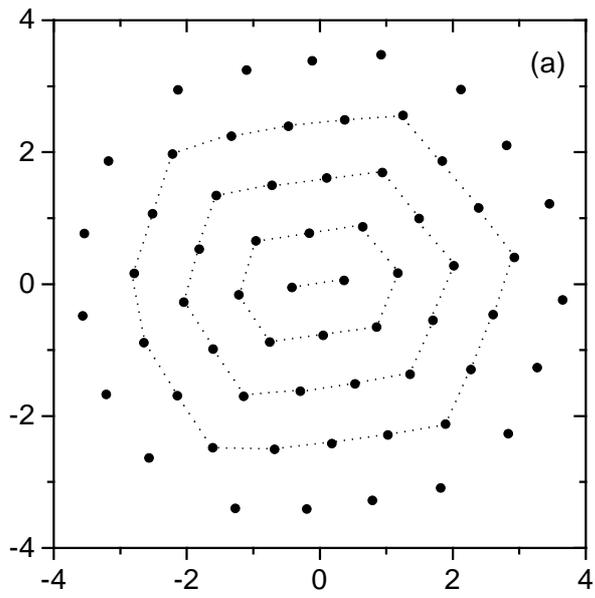
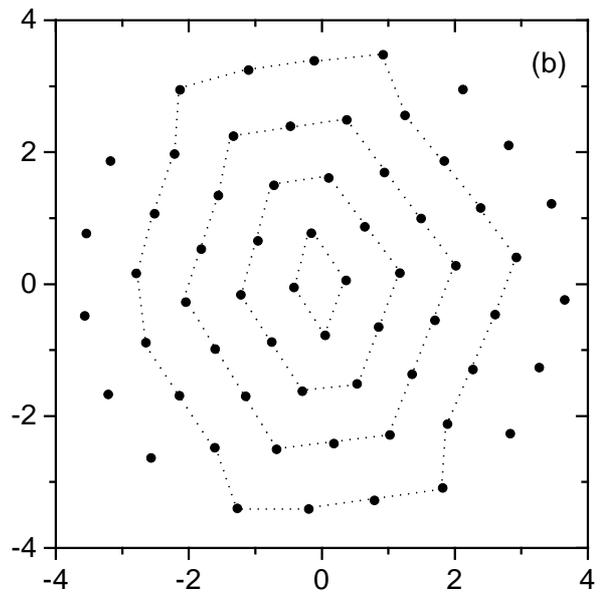
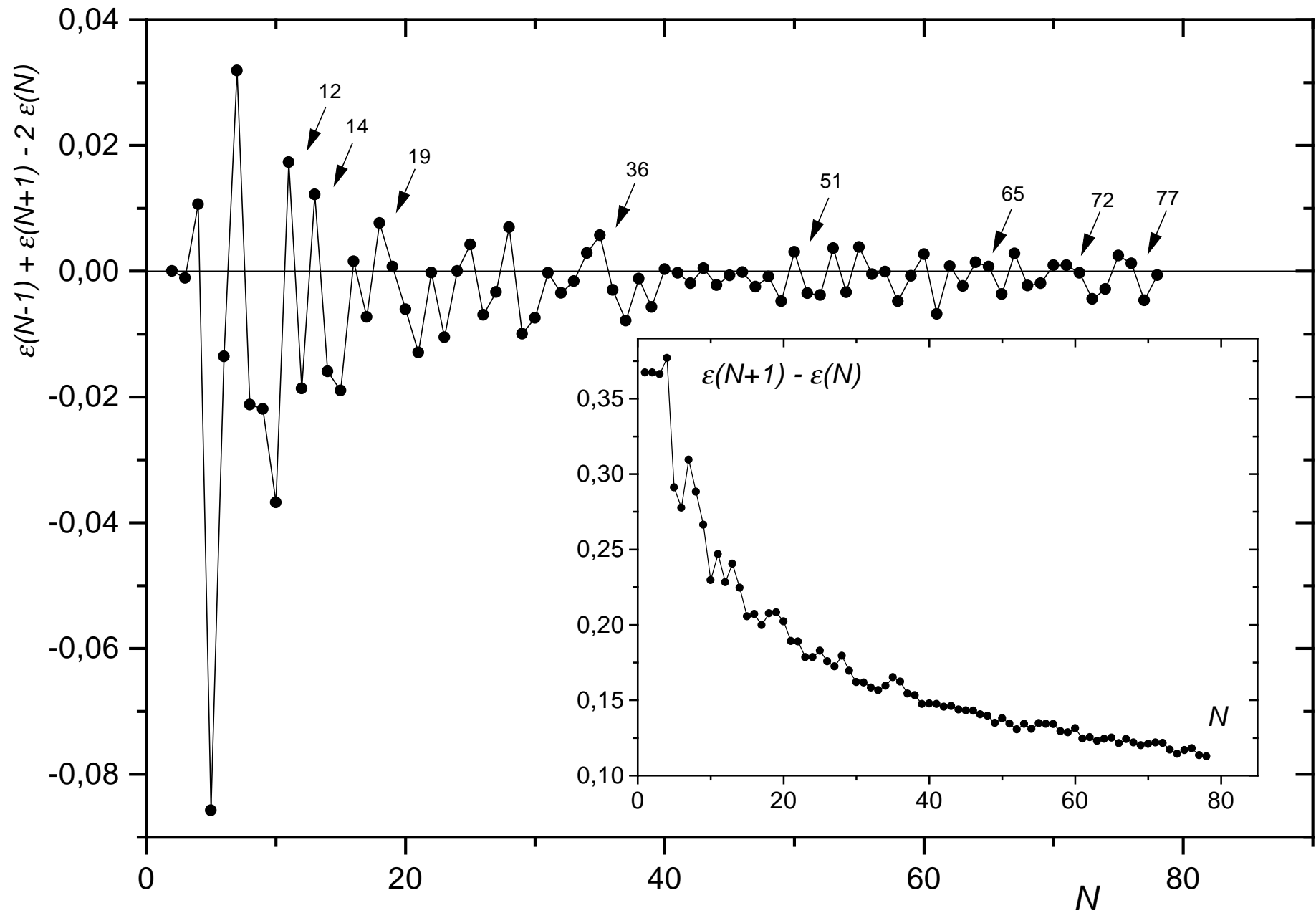
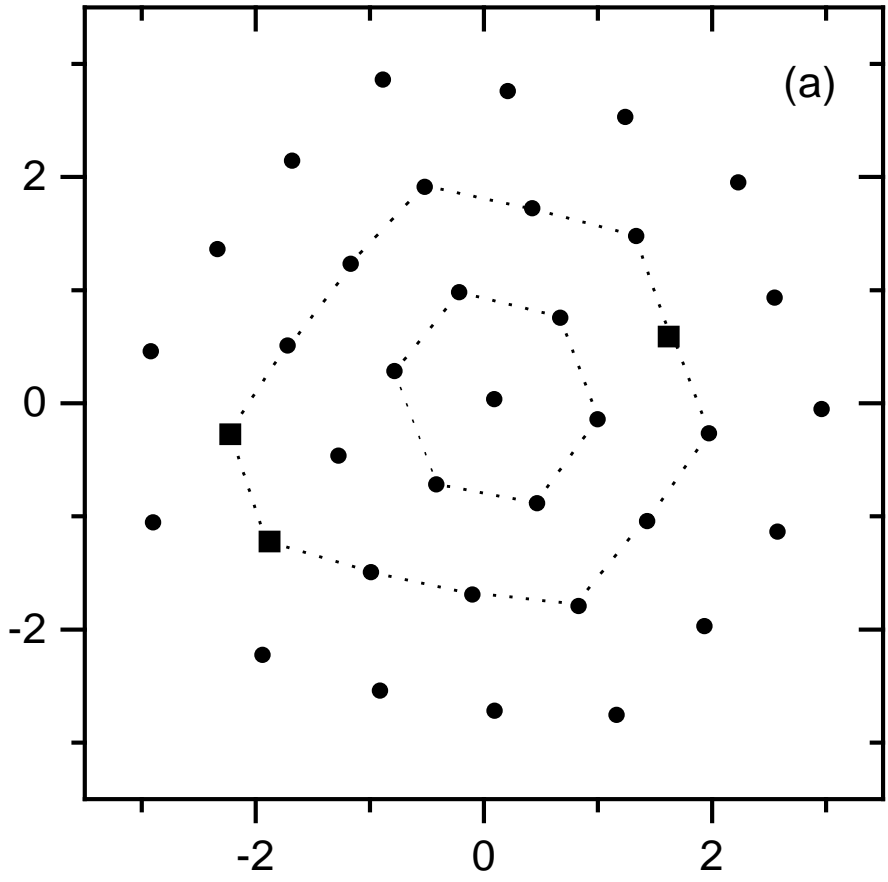
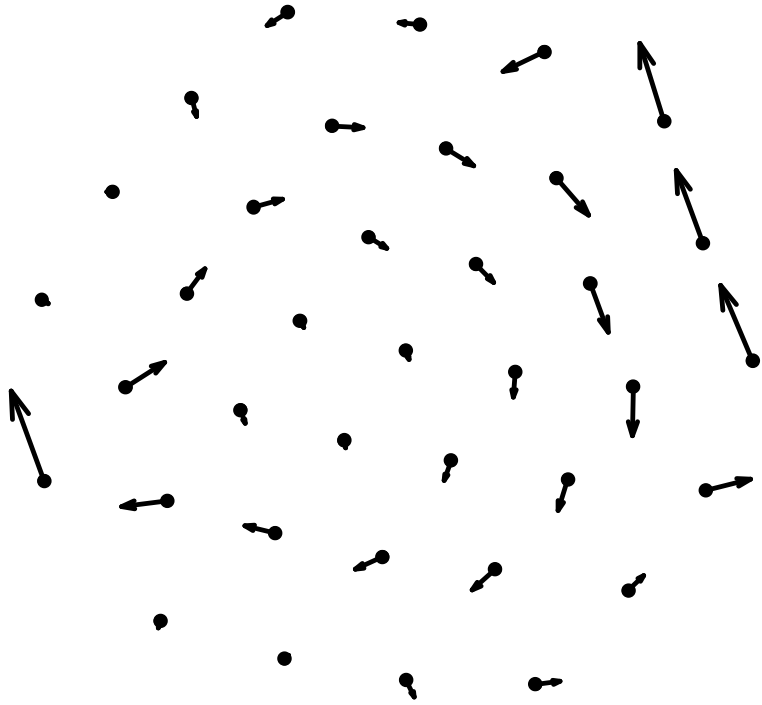
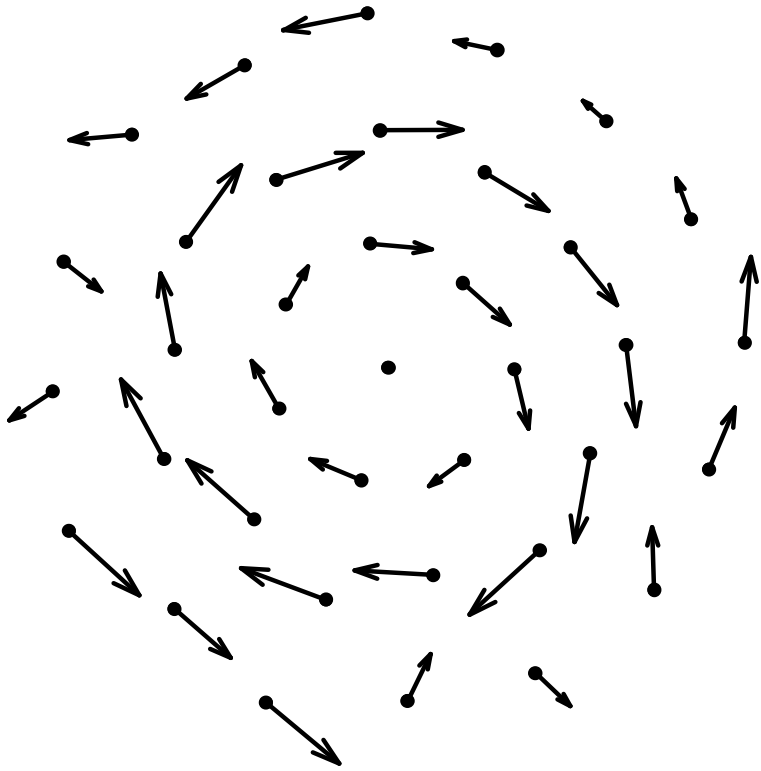
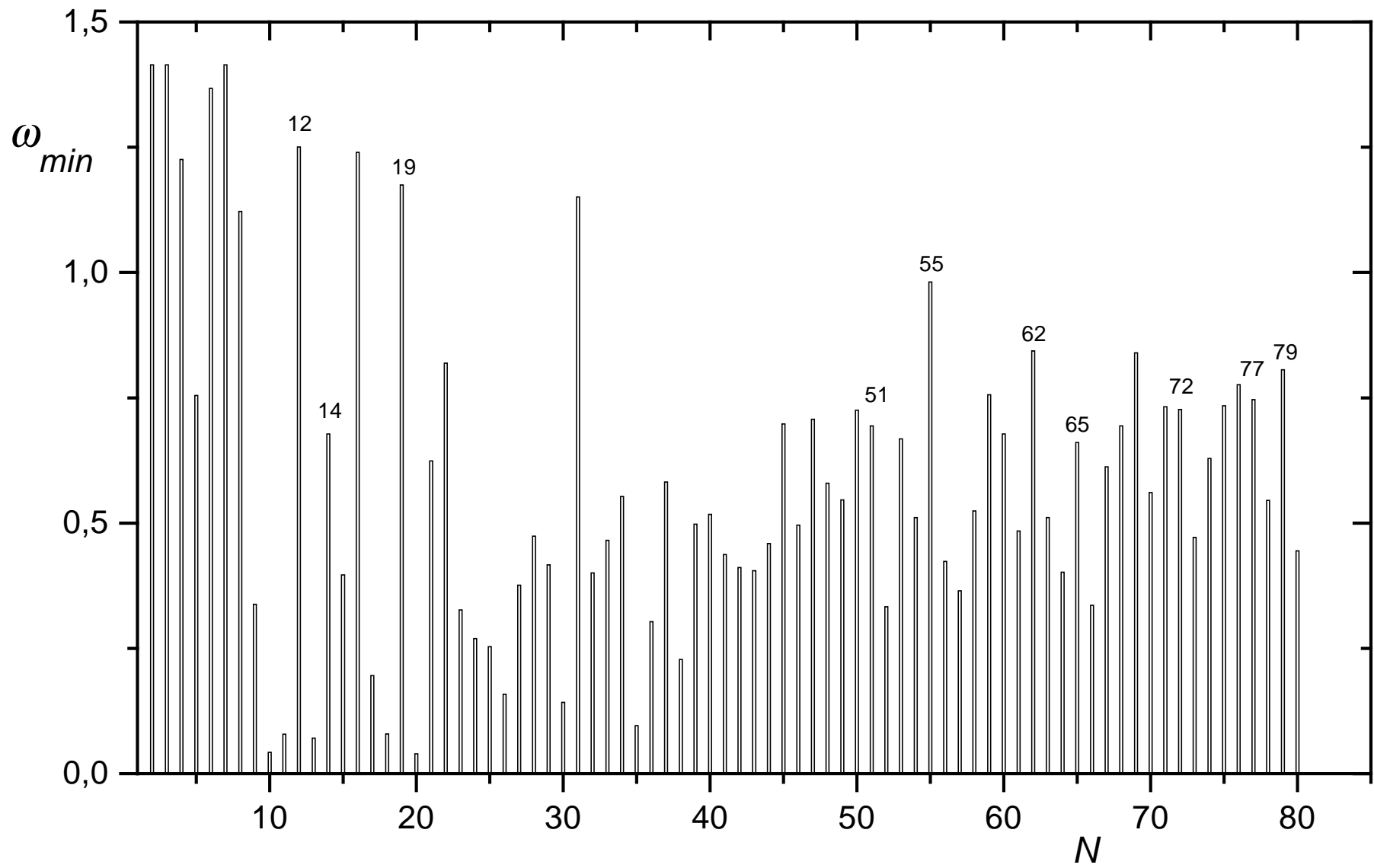
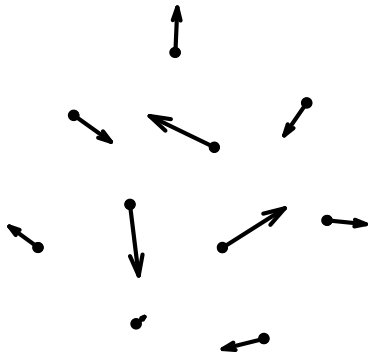
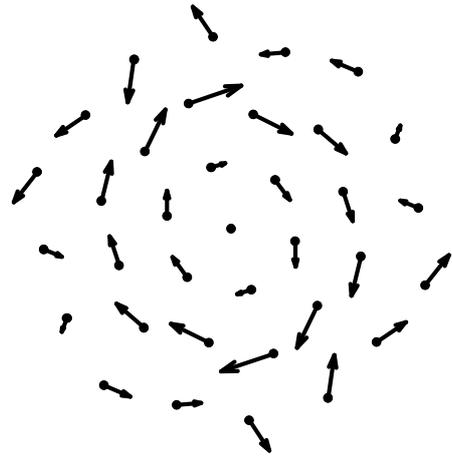
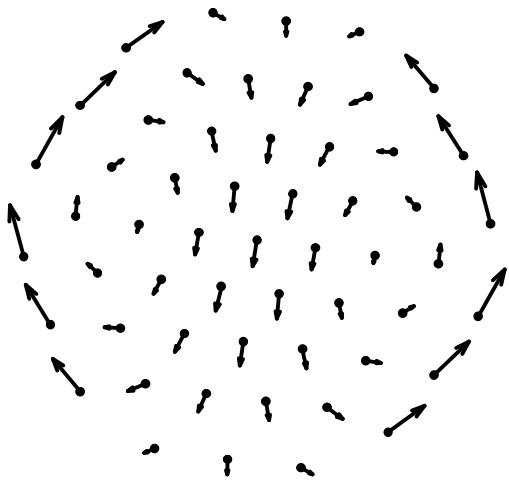
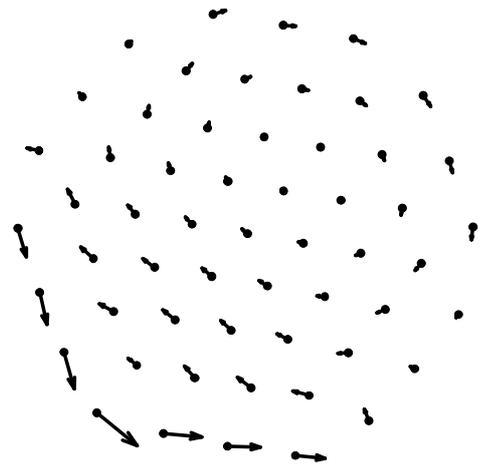
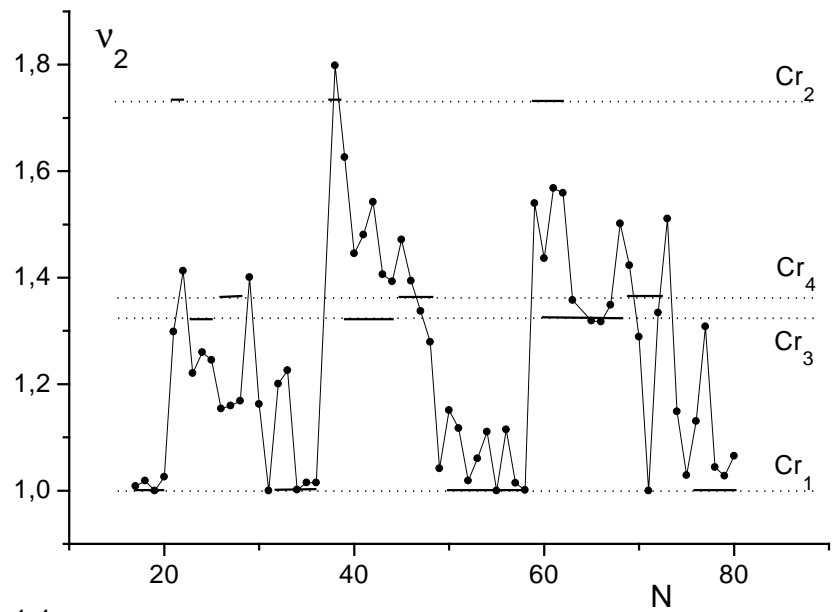
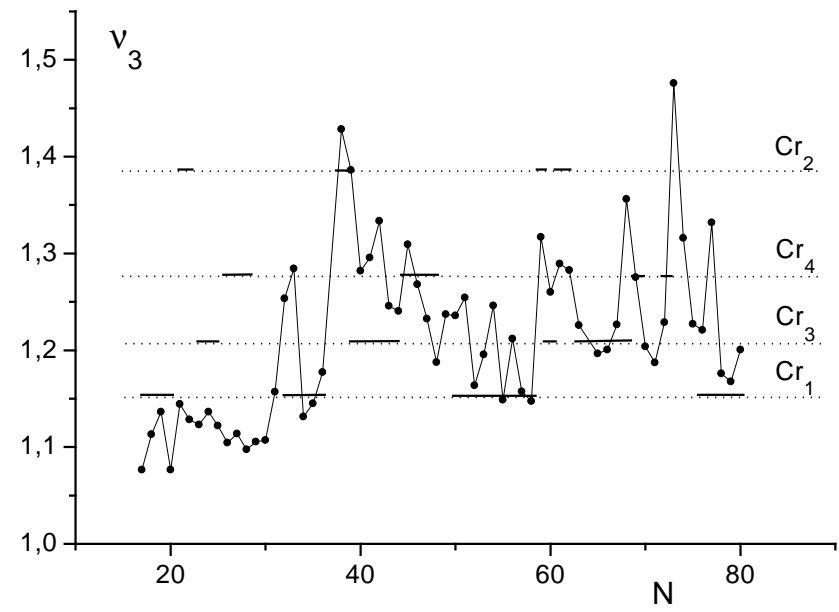
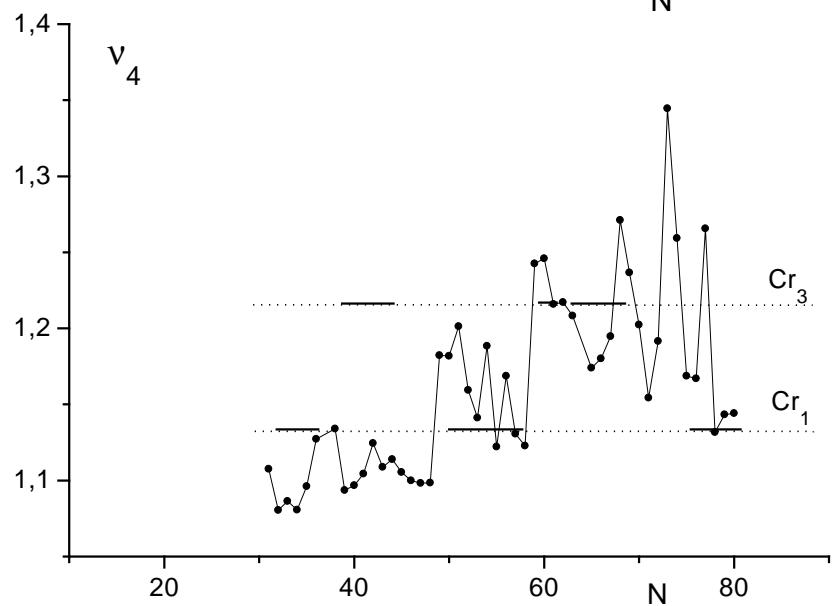
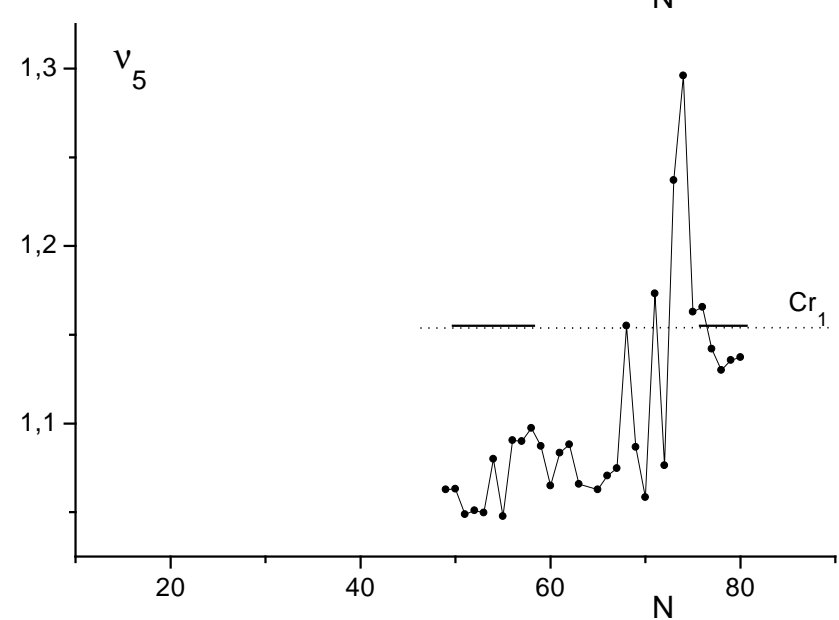
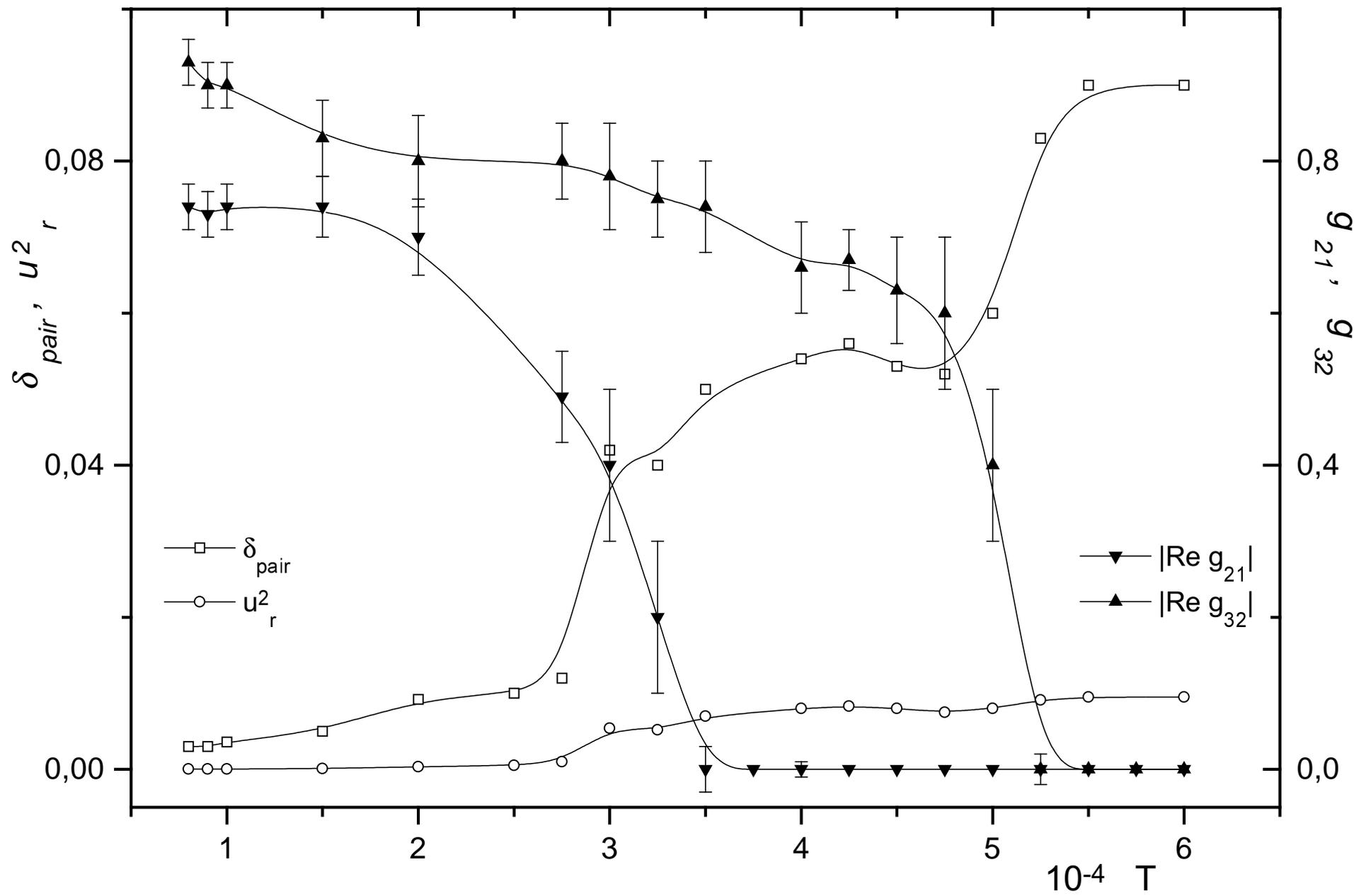
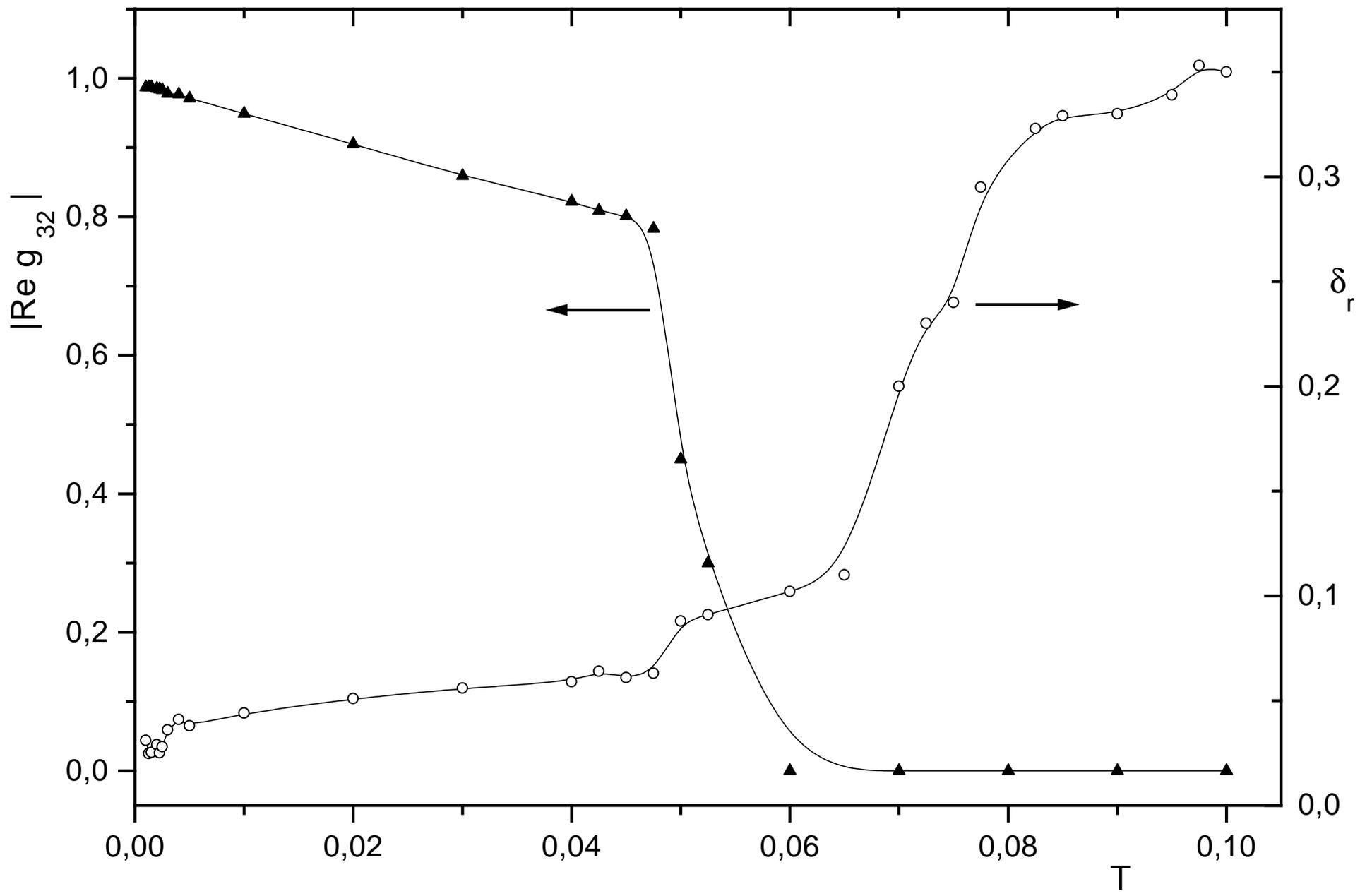
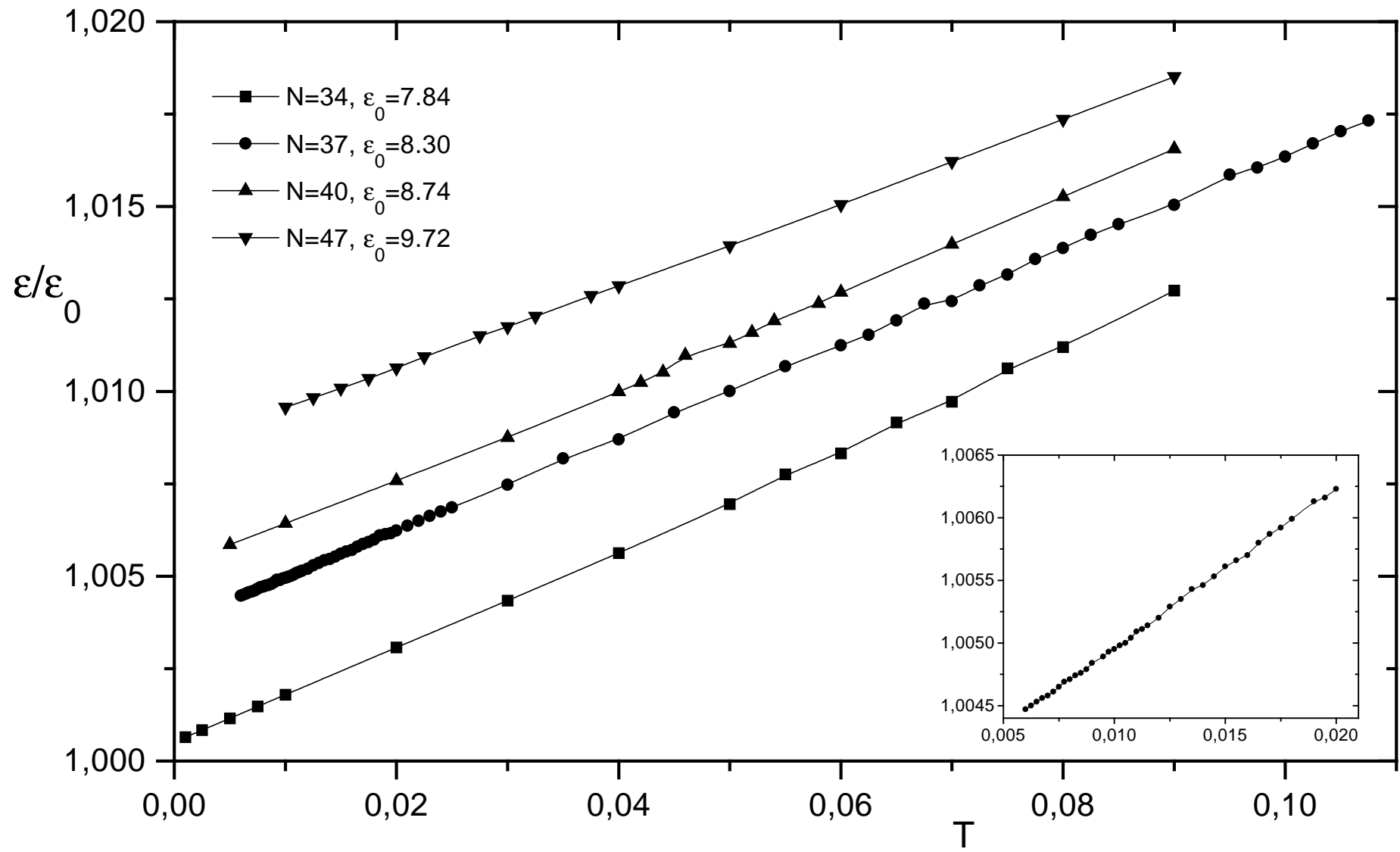
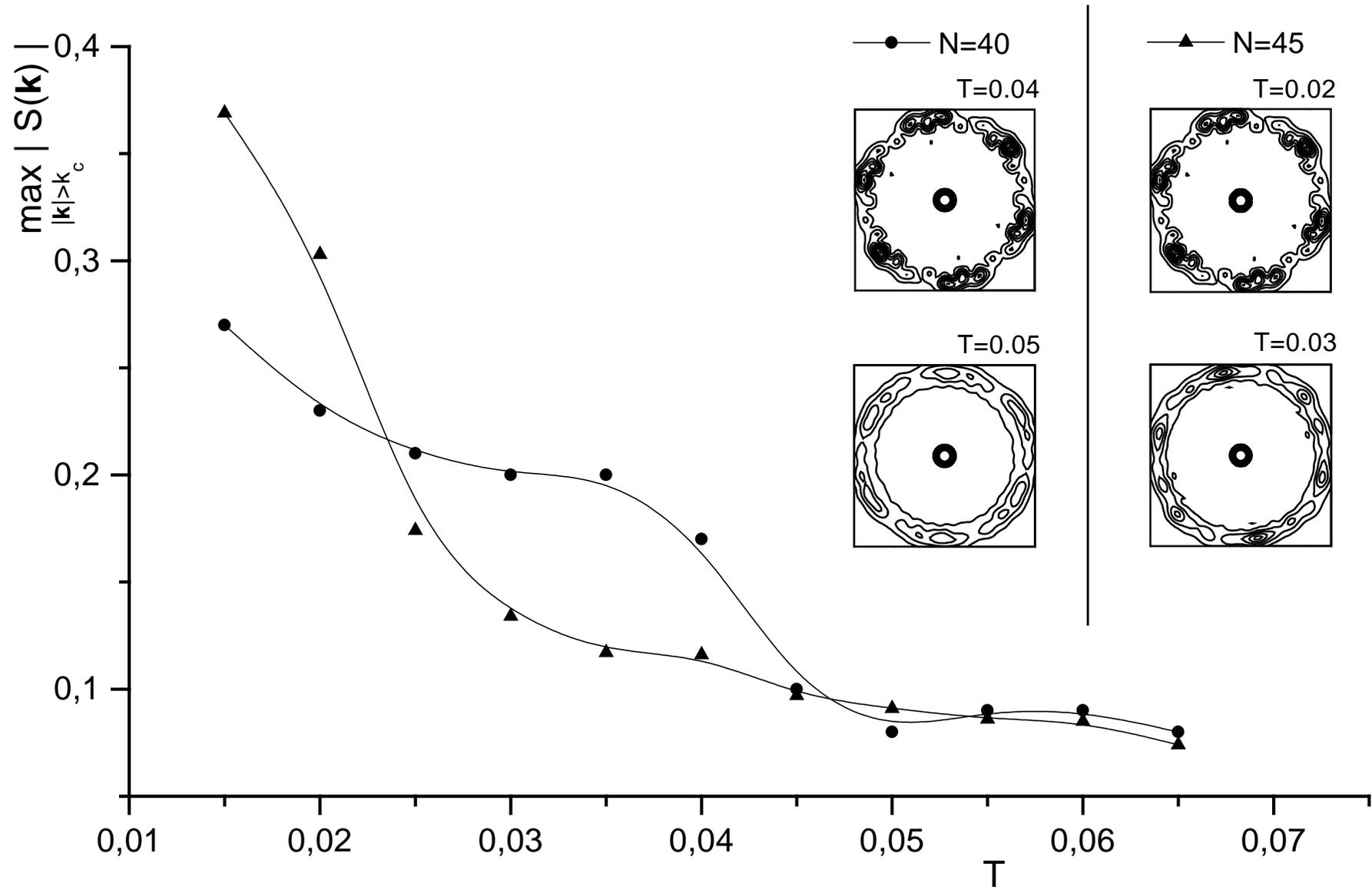
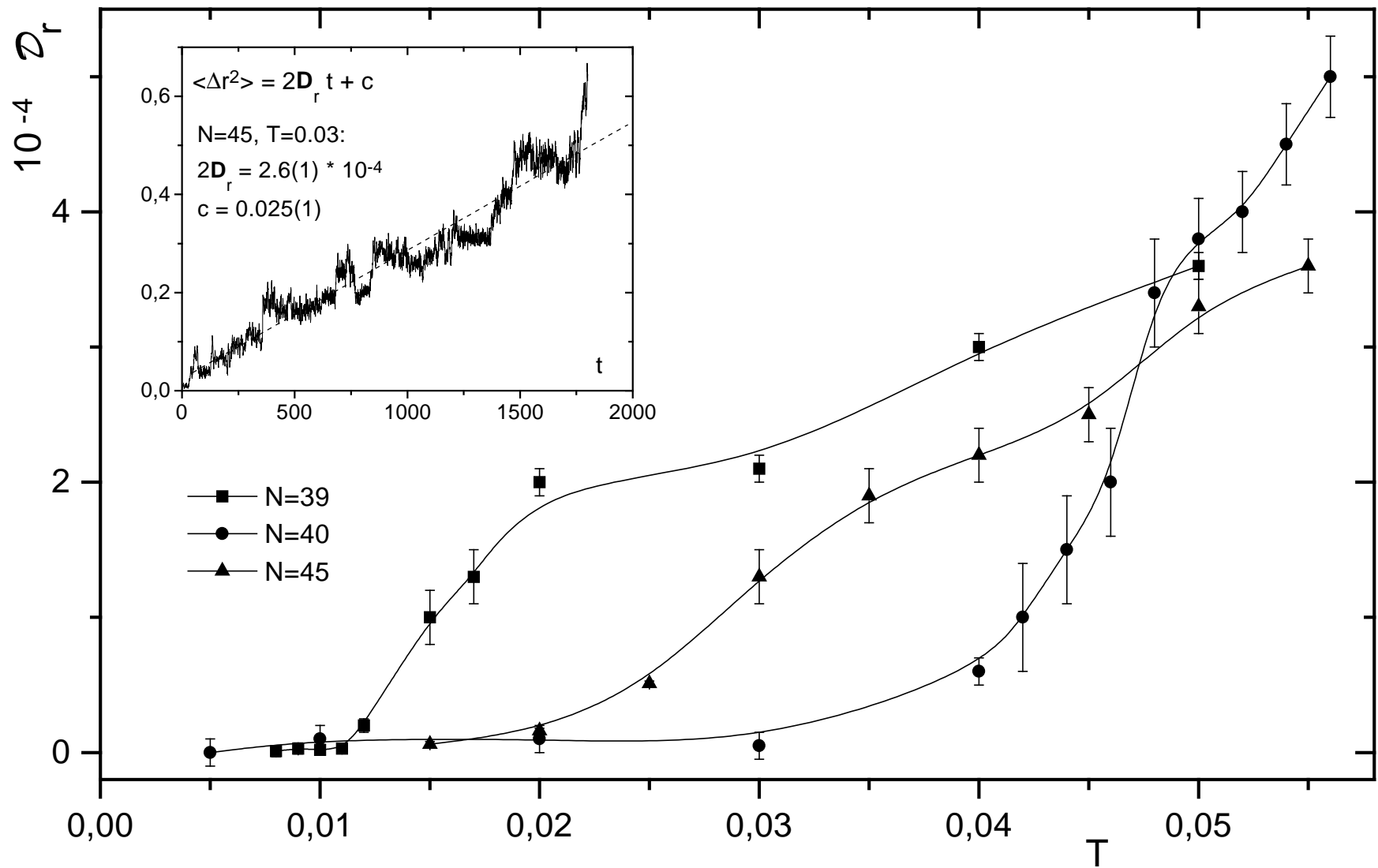
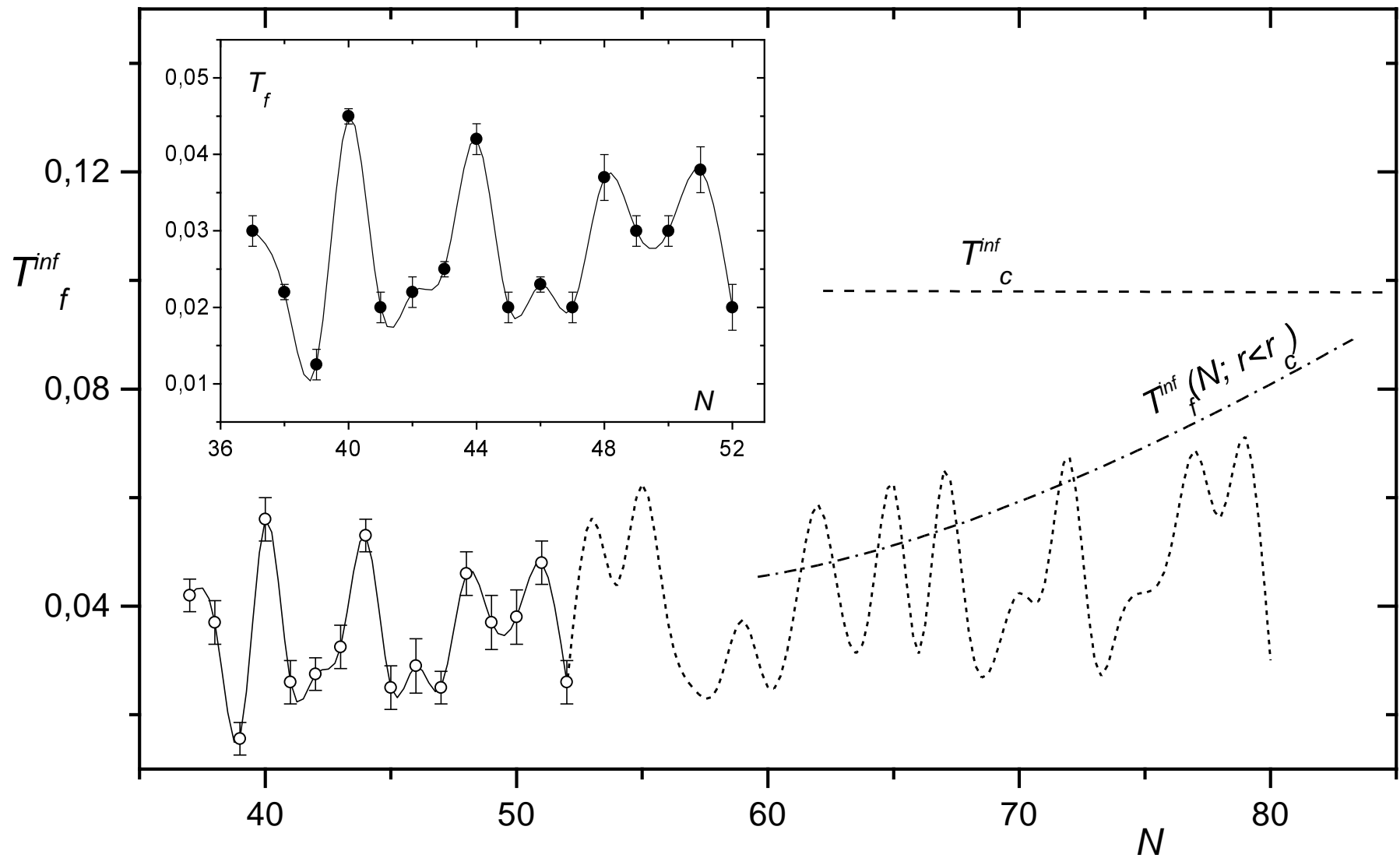
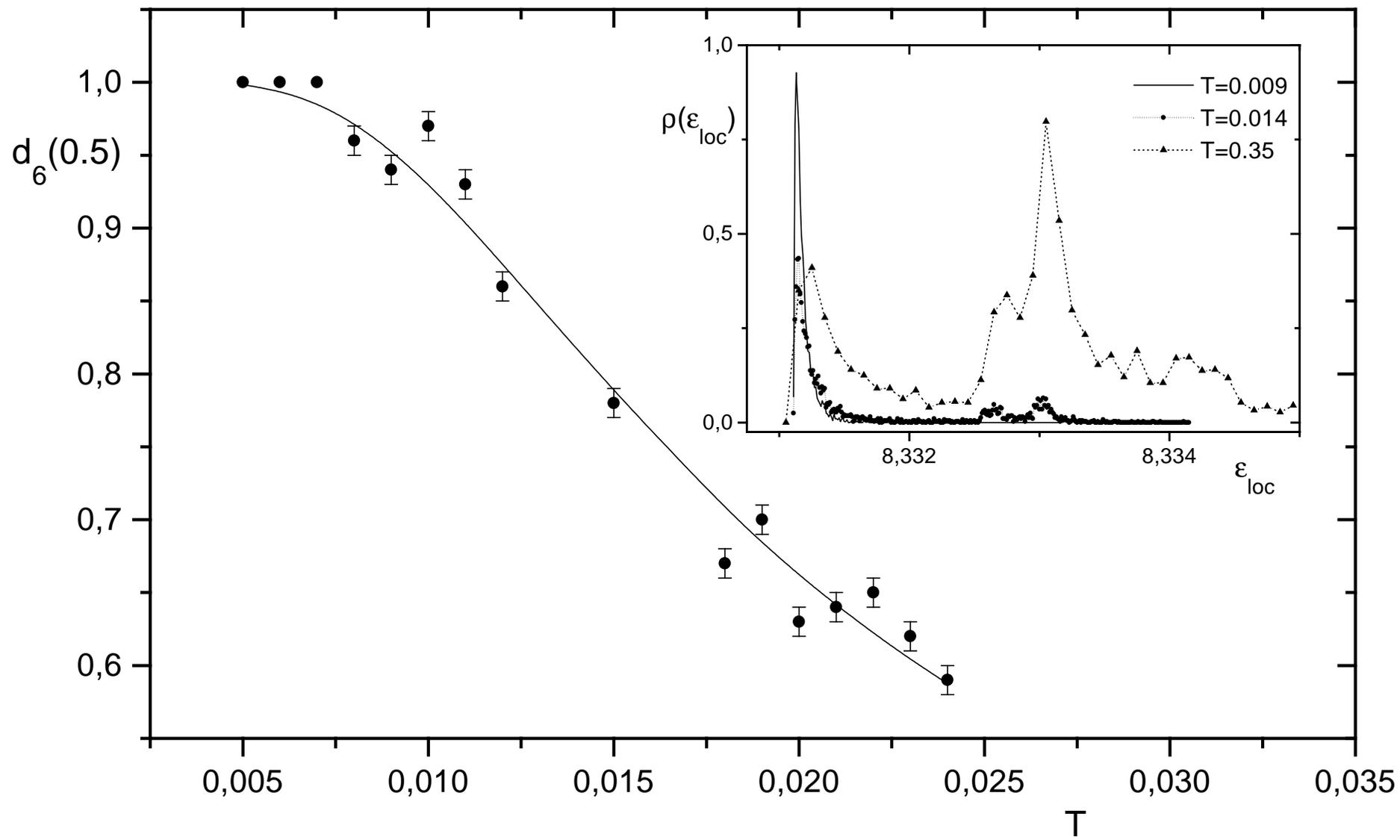
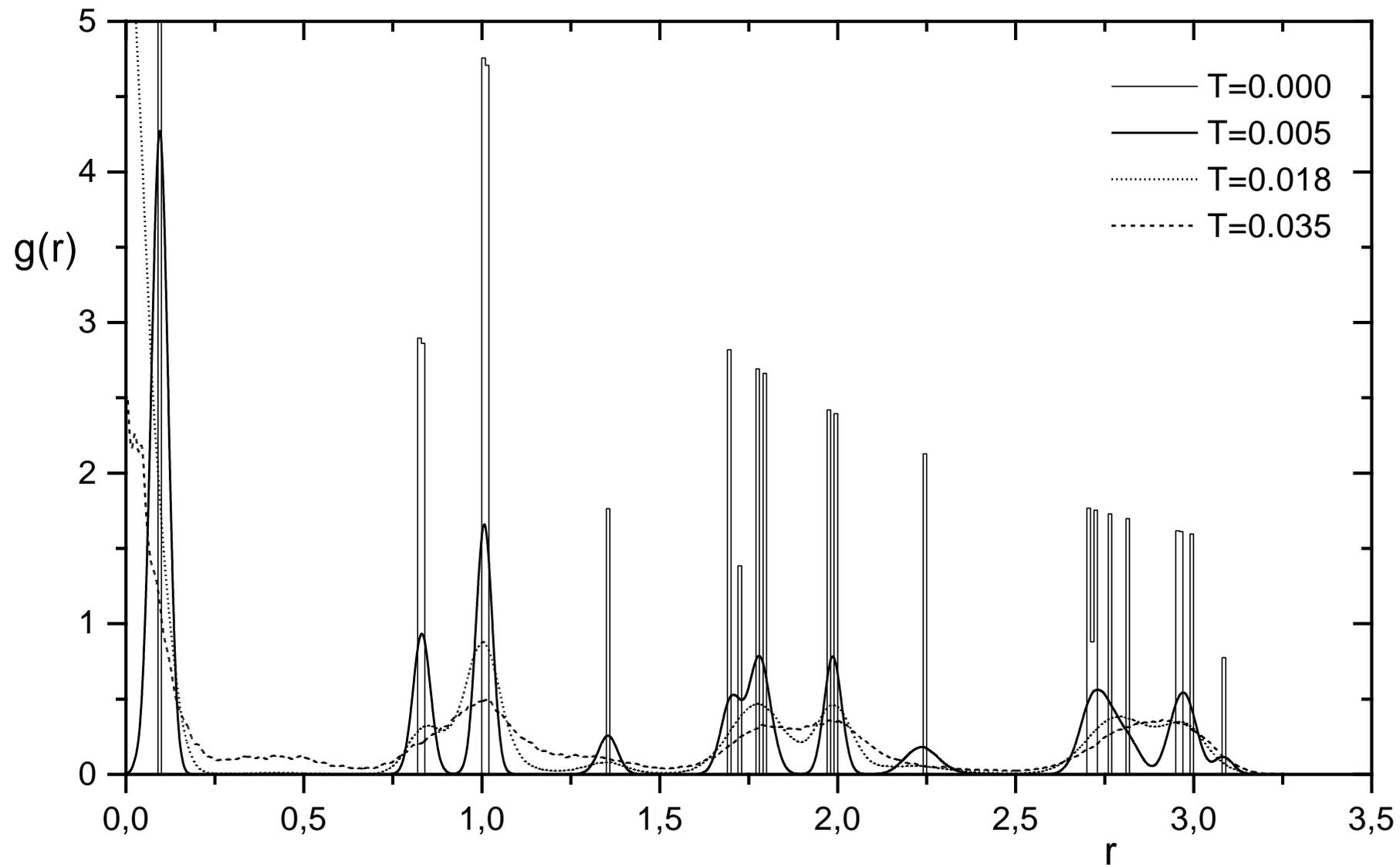
\end{document}